\documentclass[journal,11pt,doublespace,onecolumn,romanappendices]{IEEEtran}

\usepackage{setspace}
\usepackage{amsmath}
\usepackage{graphicx,psfrag,epsf}
\usepackage{float}
\usepackage{enumerate}
\usepackage{cite}
\usepackage{url}
\usepackage{amssymb,amsfonts,mathtools}
\usepackage{bm}
\usepackage{xr-hyper}
\usepackage{hyperref}
\usepackage{cleveref}
\usepackage{color, colortbl}
\usepackage{balance}
\usepackage{multirow}
\usepackage{wrapfig,bbm}
\usepackage{diagbox}
\usepackage{pdfpages}
\usepackage{booktabs}
\usepackage{float}
\usepackage{caption}
\usepackage{authblk}

\makeatletter
\def\th@plain{%
  \thm@notefont{}
  \itshape 
}
\def\th@definition{%
  \thm@notefont{}
  \normalfont 
}
\makeatother

\DeclarePairedDelimiterX{\norm}[1]{\lVert}{\rVert}{#1}
\DeclarePairedDelimiterX{\bnorm}[1]{\biggl\lVert}{\biggr\rVert}{#1}
\DeclarePairedDelimiterX{\abs}[1]{\lvert}{\rvert}{#1}

\newtheorem{definition}{Definition}

\newtheorem{theorem}{Theorem}

\newtheorem{example}{Example}
\newtheorem{assumption}{Assumption}

\def\E{\mathbb{E}} 
\def\P{\mathbb{P}} 
\def\T{{ \mathrm{\scriptscriptstyle T} }} 
\def\i{\mathbbm{1}} 
\def\v{\bm{1}} 
\def\z{\bm{0}} 
\def\w{\bm{w}} 
\def\u{\bm{u}} 
\def\x{x} 
\def\y{y} 
\def\diag{\textrm{diag}}
\def\tr{\textrm{tr}}
\def\red{\textrm{red}}
\def\green{\textrm{green}}
\def\blue{\textrm{blue}}
\def\black{\textrm{black}}
\def\de{\overset{\Delta}{=}}

\bmdefine{\bX}{X}
\bmdefine{\bv}{v}
\bmdefine{\bA}{A}
\bmdefine{\bgamma}{\gamma}
\bmdefine{\btheta}{\theta}
\bmdefine{\balpha}{\alpha}

\newcommand{\RNum}[1]{\uppercase\expandafter{\romannumeral #1\relax}}

\makeatletter
\newcommand*{\indep}{%
  \mathbin{%
    \mathpalette{\@indep}{}%
  }%
}
\newcommand*{\nindep}{%
  \mathbin{
    \mathpalette{\@indep}{/}%
  }%
}
\newcommand*{\@indep}[2]{%
  \sbox0{$#1\perp\m@th$}
  \sbox2{$#1=$}
  \sbox4{$#1\vcenter{}$}
  \rlap{\copy0}
  \dimen@=\dimexpr\ht2-\ht4-.2pt\relax
  \kern\dimen@
  \ifx\\#2\\%
  \else
    \hbox to \wd2{\hss$#1#2\m@th$\hss}%
    \kern-\wd2 %
  \fi
  \kern\dimen@
  \copy0 
}
\makeatother

\makeatletter
\newcommand\Autoref[1]{\@first@ref#1,@}
\def\@throw@dot#1.#2@{#1}
\def\@set@refname#1{
    \edef\@tmp{\getrefbykeydefault{#1}{anchor}{}}%
    \xdef\@tmp{\expandafter\@throw@dot\@tmp.@}%
    \ltx@IfUndefined{\@tmp autorefnameplural}%
         {\def\@refname{\@nameuse{\@tmp autorefname}s}}%
         {\def\@refname{\@nameuse{\@tmp autorefnameplural}}}%
}
\def\@first@ref#1,#2{%
  \ifx#2@\autoref{#1}\let\@nextref\@gobble
  \else%
    \@set@refname{#1}
    \@refname~\ref{#1}
    \let\@nextref\@next@ref
  \fi%
  \@nextref#2%
}
\def\@next@ref#1,#2{%
   \ifx#2@ and~\ref{#1}\let\@nextref\@gobble
   \else, \ref{#1}
   \fi%
   \@nextref#2%
}
\makeatother

\makeatletter
\let\oldtheequation\theequation
\renewcommand\tagform@[1]{\maketag@@@{\ignorespaces#1\unskip\@@italiccorr}}
\renewcommand\theequation{(\oldtheequation)}
\makeatother

\makeatletter
\newcommand*{\addFileDependency}[1]{
  \typeout{(#1)}
  \@addtofilelist{#1}
  \IfFileExists{#1}{}{\typeout{No file #1.}}
}
\makeatother

\begin{document}

\title{Subset Privacy: Draw from an Obfuscated Urn}

\author{Ganghua~Wang and Jie~Ding 
\thanks{G.~Wang and J.~Ding are with the School of Statistics, University of Minnesota, Minneapolis, Minnesota 55455, USA.
}
\thanks{This research was funded by the Army Research Office (ARO) under grant number W911NF-20-1-0222, and the GIA award from the Office of the Vice President for Research, University of Minnesota.}
}


\maketitle

\begin{abstract}
With the rapidly increasing ability to collect and analyze personal data, data privacy becomes an emerging concern. In this work, we develop a new statistical notion of local privacy to protect each categorical data that will be collected by untrusted entities. The proposed solution, named subset privacy, privatizes the original data value by replacing it with a random subset containing that value. 
We develop methods for the estimation of distribution functions and independence testing from subset-private data with theoretical guarantees. We also study different mechanisms to realize the subset privacy and evaluation metrics to quantify the amount of privacy in practice. Experimental results on both simulated and real-world datasets demonstrate the encouraging performance of the developed concepts and methods. 
\end{abstract}


\begin{IEEEkeywords}
Categorical data, Data collection, Estimation, Local privacy, 
Set observation.
\end{IEEEkeywords}


\section{Introduction}
\label{sec_intro}
\IEEEPARstart{W}{ith} the rapid development of the capability to collect, store, and analyze data, an increasing concern is how to protect users' sensitive data. This privacy concern appears in every aspect of our daily lives, e.g., medical measurements, political opinions, and financial activities. For example, the public profile on social media may be abused to infer individual information for personalized advertisements~\cite{isaak2018user}. Commercial companies may collect individuals' sensitive data to analyze user behavior and enhance their product quality~\cite{Google,Facebook}. To tackle the challenges of data privacy, an increasing number of privacy-preserving methods have been recently studied.

This paper introduces a new method to provide local privacy through a specialized data collecting process. Here, the term `local privacy' concerns the scenarios that involve two parties, namely individuals who have their original data and a data collector.
The local data privacy mechanism does not require a trusted third party to store all the original data in a center, which is different from the so-called database privacy~\cite{dwork2006calibrating,chaudhuri2011differentially,dwork2014algorithmic,avella2019privacy,awan2020structure}. 

Various notions of local privacy have been developed to ensure privacy-aware data collection and learning. 
Rooted in information theory, the notion of information-theoretic security combines cryptographic mechanisms with channel coding techniques to guarantee that the
transmitted messages cannot be decoded from eavesdropping, even by an adversary with unlimited computing power~\cite{bloch2008wireless,liang2009information}.
For example, a popular cryptographic solution to perform secure computation is secure multi-party computing~\cite{yao1982protocols,chaum1988multiparty}, which has received renewed interest for its recent use in federated learning~\cite{kaissis2020secure}.
To privatize the identity of individuals and also the value of the original data, a popular approach to evaluating local privacy is local differential privacy~\cite{evfimievski2003limiting,duchi2013local,erlingsson2014rappor,sarwate2014rate,bassily2015local}. Two common techniques are adding noises for continuous data~\cite{dwork2006calibrating,awan2020structure}, and randomized responses for categorical data~\cite{warner1965randomized,qin2016heavy,cormode2018privacy,yang2020local}.
To release a dataset that privatizes individual identities, the $k$-anonymity method and its variants such as $l$-diversity and $t$-closest~\cite{bayardo2005data, machanavajjhala2007diversity, li2007t} use transformations to ensure that any individual is indistinguishable from at least $k$-1 other individuals in the database. 
Two significant limitations of $k$-anonymity are its dependence on some quasi-identifiers assumed to be related to the subject identification and scalability~\cite{aggarwal2005k}. 
Our general goal of local privacy is to privatize individual-level data while enabling the possibility to uncover population-level information. 
We will focus on categorical data that may be naturally generated or discretized from continuous alphabets.
What we develop in this work provides a different perspective from the methods above. 
The main idea is to collect a subset containing the original value instead of perturbing that value. In this way, the individuals can obfuscate the private data before the collection occurs. The corresponding privacy is named \textit{subset privacy}. 
An important aspect that distinguishes subset privacy from other statistical privacy methods is that the collected data do not distort the original data. To our best knowledge, this is the first probabilistic approach that faithfully retains the real message. The noise-adding mechanism and randomized response technique in the local differential privacy will change the true value with some probability. In some application domains, such faithfulness is essential for the data collector who will not publish misleading data. 

Another desirable aspect of subset privacy is its user-friendly interface during data collection. It never requires the individuals to input the original value into the collection system, such as a questionnaire. In particular, we will introduce a simple realization of subset privacy, where an individual only needs to answer whether the true value is in a subset or not. 
Also, the collection process will encourage faithful reporting of information since the data obfuscation is transparent to individuals.
We will show that data analysts can still obtain accurate population-level information such as the distribution and statistical relationships. 

    \begin{example}[Obfuscated urn]
        \label{toy_example}
        Suppose that we have an urn that contains $p=4$ kinds of colored balls, say black, red, green, and blue. In an experiment, each of the $n$ individuals draws a ball from this urn and then puts it back. The procedure generates $n$ original data values. Due to privacy, the data collector asks each individual whether a (randomly-generated) subset (e.g., $\{\red, \blue\}$) includes the actual color. Assuming everyone's answer is honest, our observation consists of two-element sets, each containing an individual's actual color. 
    \end{example}

    The main contributions of this work are summarized below. 
First, we develop a new local data privacy methodology for categorical data. 
We provide several practical mechanisms to realize subset observations, and methods to quantify the amount of privacy leakage with different perspectives and interpretations. 
We also discuss its difference with (local) differential privacy at the end of the paper. 
Second, we develop parameter estimation methods under subset privacy, including the maximum likelihood estimation and moment-based estimation that is more computationally efficient. We provide asymptotic analysis for both estimation methods. Some extensions to high-dimensional cases where $p$ grows with $n$ are also discussed. 
Finally, we study the contingency table-based independence testing under the subset privacy. We show that some classical testing methods can be adapted to tackle subset observations. 
We demonstrate the developed concepts, methods, and theories by several examples and data experiments. 
%
We also developed a comprehensive Python package ``SubsetPrivacy'' with detailed documentation and case studies (in the supplementary material). All the experimental results are reproducible from the software package. We will publish it as open-source software once the paper is published.
        
The rest of the paper is organized as follows. Section~\ref{sec_form} formulates subset privacy and its quantification. Section~\ref{sec_est} develops parameter estimation methods for the subset data.
Section~\ref{sec_2D} discusses the contingency table of two variables and develops independence tests. Section~\ref{sec_exp} demonstrates the proposed method on real-world applications.
Section~\ref{sec_con} concludes the paper with some additional remarks.
The Appendix includes proofs, extensive experimental studies on the proposed methods and theories, and further discussions on the comparison with differential privacy.

\section{Subset Privacy}
\label{sec_form}
    \textbf{Notation}.
         We let $[p]=\{1,2,\ldots,p\}$. For a set $a$, let $\abs{a}$ denote its cardinality. Let $a^c$ denote the complement of a subset $a$, and $\i$ an indicator function.
         Let $\log$, $\ln$, $0\log 0$ denote the  base-2 logarithm, natural logarithm, and zero, respectively. 
         Let $\v$, $I_p$, $\diag(\w)$, $\w^\T$, and $\w \odot \u$ denote the all one vector, $p\times p$ identity matrix, diagonal matrix expanded by an vector $\w$, the transpose of the vector $\w$, and the outer product of $\w, \u$, respectively. 
         We define $M^\dag$ as the pseudo-inverse matrix of $M$, $\tr(M) = \sum m_{i i}$ the trace of a matrix, and $\norm{\w}_q = (\sum w_i^q)^{1/q}$ the $L_q$ norm of a vector $\w$. 
         We use $\E$ for expectation and $\P$ for probability. We write $X \indep Y$ if $X $ is independent with $Y$, and $\nindep$ if dependent.

    \subsection{Formulation of subset privacy} 
    \label{subsec_formulate}
We will focus on a scalar categorical random variable $X$, and then extend it to more general cases in \Autoref{subsec_product}. 

        \textit{Raw data}: $X \in [p]$ is assumed to be a random variable with distribution $\P\left(X=j\right)=w_{j}$ for $j=1,2,\ldots, p$. We also write $X \sim p_{\w}, \w = (w_1, \ldots, w_p)^\T$, where $p_{\w}$ is the probability mass function (PMF) and $p_{\w} (j)=w_j$. 
        
        \textit{Privatized data}: $A \in \mathcal{A}$ is assumed to be a random variable whose joint distribution with $X$ exists. 
        Here, $\mathcal{A} = \{a: a \in [p]\}$ denotes the set of all the subsets of $[p]$.
        
        Recall that we will collect $A$ instead of $X$ to infer population-level information while protecting each individual's privacy. We want $\abs{A}\geq 2$ so that the true category cannot be immediately identified. Consequently, we need $p \geq 3$ for creating a non-trivial subset. For now, we assume that $p \geq 4$, and defer the special cases of $p=2,3$ to \Autoref{subsec_p23}. The case of $p=1$ is not interesting since the variable is deterministic. 
        Also, it is appealing to design such $A$ that does not immediately imply $X$. Ideally, the only information provided by $ A $ about $ X $ is the event $X \in A $, which is so-called non-informative property. 
        This motivates the following definition. 
        \begin{definition}[Subset privacy]
            \label{subset_privacy}
            A mechanism as described by a Markov chain $X \to A$ meets subset privacy, if $\abs{A} \geq 2$ and
            for any $x \in [p], a \subseteq [p]$,
            \begin{equation} \label{noninfo}
                \P_{X \mid A}(x \mid a)=\frac{ p_{\w}(x)}{\P_{\w}(x \in a)} \mathbbm{1}_{\{x \in a\}}(x) .
            \end{equation}
        \end{definition}

        Next, we introduce a general mechanism to generate a Markov chain $X\to A$ satisfying the subset privacy.
        \begin{definition}[Conditional mechanism]
            \label{cond_design}
            A {conditional mechanism} is a Markov chain $X \rightarrow A$ whose transition law is determined by 
            \begin{align*}
                P(A=a \mid X=j) = \mu_a\i_{j \in a},
                \quad \forall a \subseteq[p], \, j \in [p], 
            \end{align*}
            where $\mu_a$ satisfies
            $\sum\limits_{a:j \in a} \mu_a = 1, \forall j \in [p]$, and $\mu_a=0$ if $|a|<2$.
            
            Any specific realization $\{ \mu_a, a \in \mathcal{A} \}$ from a conditional mechanism is referred to as a \textit{conditional design}.
            We will use $\{ \mu_a,  a \in \mathcal{A} \}$ to represent a conditional design in the paper. 
        \end{definition}      
        
        We will show after Definition~\ref{ind_design} that there exist infinitely many conditional designs (for $p \geq 4$). 
        It can be verified that the conditional mechanism is the only mechanism to meet \Autoref{subset_privacy}. Nevertheless, it is nontrivial to find all the conditional designs that meet the constraints $\sum\limits_{a:j \in a} \mu_a = 1, \forall j \in [p]$. Additionally, such a mechanism requires the data collector to know the true value $X$ to realize the design and generate $A$. As a result, the mechanism is often convenient only from the \textit{data publisher's perspective}. Because of the above reasons, we will introduce a special conditional mechanism named independent mechanism that is easy to implement. 
        The idea behind the independent mechanism is to generate a Markov chain $X \to A$ through another chain $[X,\tilde{A}] \rightarrow A$, where $\tilde{A}$ is independent with $X$, and $A$ satisfies 
        $$ 
        A = \tilde{A} \quad \textrm{if } X \in \tilde{A}, \quad 
        \textit{ and }  \quad  A =\tilde{A}^c \quad \textrm{if } X \not\in \tilde{A}.
        $$ 
        In other words, the mechanism independently generates a subset $\tilde{A}$ and then uses it or its complement as the subset observation $A$. 
        \begin{definition}[Independent mechanism]
            \label{ind_design}
            An independent mechanism is a Markov chain
            $X \rightarrow A$ induced by another Markov chain $[X,\tilde{A}] \rightarrow A$,
            whose transition law is given by 
            \begin{align*}
                & \tilde{A} \indep X, \ \P(\tilde{A}=\tilde{a}) = \nu_{\tilde{a}}, \\
                &\P(A=a \mid X=j, \tilde{A}=\tilde{a}) = \i_{j \in \tilde{a}, a=\tilde{a}} + \i_{j \notin \tilde{a}, a=\tilde{a}^c} , \quad \forall a,\tilde{a} \subseteq[p], \, j \in [p],
            \end{align*}
            where $\nu_a$ satisfies
            $\sum\limits_{a \in \mathcal{A}} \nu_a = 1$, and $\nu_a=0$ if $|a|=1$ or $|a|=p-1$.
            
            We will use $\{ \nu_a, a \in \mathcal{A} \}$ to represent an independent mechanism throughout the paper.
            In practice, in order to avoid the trivial case that $a=[p]$, we usually require $\nu_{\{[p]\}} = \nu_{\{\emptyset\}} = 0$.
        \end{definition}  
            It is conceivable that the final observation $A$ will be non-informative, since $A$ is constructed based on $\tilde{A}$ that is independent with $X$. Technically, the transition law of $X \to A$ in \Autoref{ind_design} can be characterized as $P(A=a|X=j) = (\nu_a+\nu_{a^c})\i_{j \in a}$, and thus every independent design $\{ \nu_a,  a \in \mathcal{A} \}$ leads to a conditional design $\{\mu_a = \nu_a+\nu_{a^c},  a \in \mathcal{A} \}$. In practice, the data collector only has to ask an individual whether the true value is in a subset $\tilde{A}$ or not, hence the independent mechanism is most useful for data collection. 
            Also, we note that there exist infinitely many independent designs that correspond to the same conditional design, but not every conditional design corresponds to an independent design. 
            
        As an example, a specific design under the independence mechanism is assigning all allowed subsets with the same probability.
        \begin{definition}[Uniform independence design]
            An independence design 
            $\{ \nu_a,  a \in \mathcal{A} \}$ is uniform if  $$\nu_a = \begin{cases}0, \quad &\textrm{ if } |a|=0,1,p-1,p \\ \frac{1}{2^p-2p-2}, \quad &\textrm{ otherwise } \end{cases}.$$
        \end{definition}   

        Before we conclude this subsection, we introduce another notation of $a$ that will be frequently used in the remainder of the paper.
        We will represent $a$ with $\v_a$, which is a vector in $\{0, 1\}^p$, with $j$-th coordinate one if $j \in a$ and zero otherwise. Let $\v_A$ be the  random vector corresponding to $\v_a$. 
        With the above notation, under a conditional design, the joint distribution of $X, A$ is $\P(X=j,A=a) = w_j\mu_a \i_{j \in a}$, with marginal distribution $\P(A=a)=\v_a^\T\w\mu_a$.

    \subsection{Quantification of privacy} \label{subsec_quantity}
        In this subsection, we will quantify the amount of privacy for a conditional design or a particular user.
        
        \subsubsection{Threat model}
        A potential adversary can access subset observations $\{A_1, \dots, A_n\}$. We assume that the adversary does not know the identity of the individual associated with any subset $A$ before observing the data. Its goal is to infer the individual's identity, the underlying value $X$, or both. 
        The above setting is commonly seen in practice. For example, for data publishing, entities will often first anonymize the data by removing all identifiers. 
        We allow the potential adversaries to know all or part of individuals' information. 
        
        \subsubsection{Subset-size based privacy coverage and leakage}
         It is conceivable that a `larger' subset $A$ leads to more difficulty in guessing $X$ since we would have less knowledge about $X$. Such knowledge means not only the cardinality of $A$ but also the probability of $X$ belonging to $A$. The above intuition motivates the following notion of privacy. 
            
\begin{definition}[Size privacy]
The size of an individual subset $a$ is defined as $L(a)=P(X \in a)=\v_a^\T \w$.  The \textit{size privacy coverage} of a given design is
$\tau = \mathbb{E} ( L(A) ) = \E ( \v_A^\T \w ),$
and the \textit{size privacy leakage} is
$S = 1-\tau = \E (1- \v_A^\T \w)$.
\end{definition}    

In the above definition, we used the sum of probabilities of the categories observed in a subset to measure each individual's privacy. Size privacy coverage is the average subset size. 
The defined privacy coverage and leakage are naturally interpretable for the unique subset data format. For an individual being collected the subset $a$, the same subset may also be collected from $L(a)$ portion of the population. The larger $L(a)$ is, the more ambiguity of identifying a specific individual. The size of a subset $a$ or the design is related to the underlying distribution $\w$, which is usually unknown in practice. To \textit{actively control} the lower bound of the privacy coverage, we will propose a modification of the subset privacy in Appendix~\ref{sec:dummy}. 

            \begin{example}[Obfuscated urn, continued]
                Assume that the underlying distribution is given by  $\P(\black)=0.01$, $\P(\red)=0.1$,  $\P(\green)=0.2$, and $\P(\blue)=0.69$. Suppose that we observe $a = \{ \red, \green, \blue\}$ from an individual. Then we know $X$ is not $\black$. However, this subset only eliminates a very small proportion of uncertainty. Knowing $a$ will not significantly help us guess $X$.
This is aligned with our defined subset size of $0.99$, interpreted in the way that the individual datum is very secure given the observation.
 
In another case with $a=\{\black, \red\}$, we know that a small-probability event occurs. In this case, knowing $a$ leaks more information about $X$. Quantitatively, the subset size of $a$ is $0.11$, meaning a relatively higher privacy leakage. If we use a uniform design, the average subset size (or privacy coverage) for this example is about $0.684$. 
           
\end{example}

             We note that subset privacy is a concept in a relative sense, comparing the information change before and after observing $A$. In a particular example, suppose that we know there is only one possible color, $\red$. Then, even if we observe $\{\red, \green\}$, the coverage is one because we do not leak any additional information. In this sense, it is considered private.

            
            
    \subsubsection{Information-theoretic privacy coverage and leakage}
       From an information-theoretic perspective, a data privacy mechanism $X \to Y$ may be treated as a Markov chain, where the privatized data $Y$ may take values in a general alphabet. A related notion of privacy is known as information privacy~\cite{du2012privacy,sun2016towards,diaz2019robustness,sun2019decentralized}, which measures privacy through information quantities. 

       The subset privacy can also be quantified from the information-theoretic view.  
        One natural idea is to measure the information contained in $A$ about $X$ using the mutual information~\cite{cover2012elements}.
        \begin{definition}[Mutual information privacy]
           For a given design, the mutual information privacy leakage is $I(X;A) = \E \log \frac{P(X, A)}{P(X)P(A)}$, and the privacy coverage is the conditional entropy $H(X\mid A)  = \E\log P(X\mid A)$.  
        \end{definition}

            Another natural idea is to measure the privacy leakage by the maximal probability (or `prediction risk') to correctly guess the label $X$ given $A$. Formally, let a random variable $Y \in [p]$ be our guess for $X$. Then, $X \to A \to Y$ forms a Markov Chain since $Y$ only depends on $A$. 
            The prediction risk is the largest probability $\P(X=Y)$ among all the conditional distributions $\P_{Y\mid A}$. 
            \begin{definition}[Prediction privacy]
                For a given design, the prediction privacy leakage and coverage are respectively $$R(X;A)=\sup_{\P_{Y\mid A}} \P(X=Y), \quad P(X;A) = 1-R(X;A).$$
            \end{definition}

    \subsection{Extension to multiple variables}\label{subsec_product}
        In this subsection, we show that the one-variable case can be generalized to the multi-variable case. 
        Suppose we have a $d$-dimensional random vector $\bX= (X_1, \ldots, X_d)^\T$, each $X_k$ being a discrete random variable of $p_k$ categories. The observation is $A$, a random variable whose value is a subset of all possible outcomes of $\bX$. We can construct a mapping $\bX \to X$, where $X$ has $p=\prod\limits_{i=1}^d p_i$ categories. In this way it degenerates to $d=1$ case. 
        
        A practical concern of constructing a mapping is a large sample space of the subset $A$, as $\abs{\mathcal{A}} = 2^p$ grows exponentially with $d$. To address this concern, we propose another way to deal with multiple variables. In particular, we restrict the form of observation $A$ to the product of subsets corresponding to each $X_k$. In other words, $A$ can be decomposed as the product of $\bA = (A_1, \ldots, A_d)$, where each $A_k$ is a subset of possible outcomes of $X_k$. For implementation, we observe $A_k$ from each $X_k$ with a conditional mechanism, then put them together as $\bA$. As a result, the conditional mechanism for multiple variables becomes the following product mechanism.
        \begin{definition}[Product mechanism]
            \label{prod_design}
             A product mechanism is a Markov chain $\bX \to \bA$ whose transition law is determined by
           \begin{align*}
                \P(\bA=\{a_1,\ldots,a_d\} \mid  \bX=\{j_1,\ldots,j_d\}) = \prod_{k=1}^d \mu_{a_k}\i_{j_k \in a_k},
                \quad \forall a_k \subseteq[p_k], \, j_k \in [p_k], k \in[d] ,
            \end{align*}
            where $\{\mu_{a_k}, a_k \subseteq[p_k]\}$ is a conditional design for each $k$.
        \end{definition}            
        
        The advantage of using a product mechanism is that the sample space of $\bA$ is a product space for every coordinate, with a cardinality $2^{\sum_k p_k}$ instead of $2^p$. It simplifies the construction of the conditional design for multiple variables and increases the parameter estimation efficiency and hypothesis testing power. As a particular case and interesting application, we will introduce the contingency table from two-variable subset data in Section~\ref{sec_2D} and demonstrate the power of the product mechanism. 

    \subsection{Extension to two- or three-category variables (\texorpdfstring{$p=2,3$}{p=2,3})} \label{subsec_p23}
        In this subsection, we propose two solutions to address the case when $p=2$ or $3$.

            The first solution is by combining every two variables with less than four categories into one variable. For example, suppose we have two binary variables, say gender (male, female) and income (low, high). We represent them as a new variable with four categories and apply subset privacy based on this new variable.  The inference of each variable is often straightforward from the inference of the combined variable. For example, the probability of observing one category is implied by the marginal distribution. The above method can be directly applied to any even number of variables.
            If we have an odd number of variables with less than four categories, a solution is to combine it with an artificial variable. 
            In a practical data collection system, the above technique is easy to implement. Specifically, the data holder (who owns the private data) and the data collector may use an open-sourced program to generate a random number as an artificial variable. 
         
     An alternative method is to introduce dummy categories to the data, which is elaborated in the Appendix~\ref{sec:dummy}. 

\section{Parameter Estimation for Subset Data}
\label{sec_est}

    In this section, we study methods to estimate the population distribution $\w$ (in \Autoref{subsec_formulate}) for a single variable $X$.
    Recall that the observed data (to the data collector/analyst) are $n$ subsets $A_1,\ldots,A_n$ generated from some design $\{ \mu_a, \forall a \in \mathcal{A} \}$ and i.i.d. original data $X_1,\ldots,X_n$. 

    \subsection{MLE and theoretical properties} 
        A standard way to estimate $\w$ is using the maximum likelihood estimator (MLE). 
        Here we discuss the model identifiability, consistency, and asymptotic normality of MLE.
        Note that $\w \in \mathcal{K}=\{\w: 0\leq w_i \leq 1, \sum\limits_{i=1}^p w_i=1\}$ has only $p-1$ degrees of freedom. We reparameterize the parameter space with 
        $$
        \Theta=\biggl\{\btheta: \btheta_i = w_i, 0\leq\btheta_i\leq1,  i=1,\ldots,p-1; \sum\limits_{i=1}^{p-1}\btheta_i\leq 1\biggr\}.
        $$ 
        Hence, we have $\w=\balpha+B\btheta$, where $\balpha=(0, \ldots, 0, 1)^\T, B=(I_{p-1}, -\v)^T$. 
        
        Let $x_i, a_i$ be $n$ i.i.d. realizations of $X, A$. 
        The log-likelihood function is  
        \begin{equation}
            \label{llh} l_n(\btheta) \de l_n(\w) = \sum_{i=1}^{n} \ln \v_{a_i}^\T (\balpha+B\btheta).
        \end{equation} 
        Since there is a one-to-one mapping between $\w$ and $\btheta$, with a slight abuse of notation, we will also write the log-likelihood function as 
        \begin{equation} \label{loglikelihood}
            l_n(\w) = \ln \P\left(A_{i}=a_{i}, i=1,2, \cdots, n\right)  
                   = \sum_{i=1}^{n} \ln \v_{a_i}^\T \w
        \end{equation} 
        (up to a constant that does not depend on $\w$).
        As we will discuss, \Autoref{loglikelihood} is more convenient for optimization, while \Autoref{llh} is suitable for theoretical analysis.
        
        First, we provide a sufficient and necessary condition to guarantee the model identifiability. 
        \begin{assumption}[Identifiability] \label{pd}
             If for all $a$ with $\mu_a > 0$, $\v_a^\T B\u=0$ holds, then $\u=\z$.
        \end{assumption}    
        \begin{theorem}\label{thm_identi}
            A subset privacy model $X \to A$ is identifiable if and only if \Autoref{pd} holds. 
            Moreover, for an independent mechanism, \autoref{pd} is equivalent to the following:  if for all $a$ with $\mu_a > 0$, $\v_a^\T \u=0$ holds, then $\u=\z$.
        \end{theorem}
        
        
        The following assumption is sufficient to guarantee the uniqueness of the MLE, which follows the proof in \cite{gentleman1994maximum}. 
        \begin{assumption}[Uniqueness of MLE]
        \label{unique}
             Matrix $RB$ is full-rank, where $R=(\v_{a_1}, \ldots, \v_{a_n})^\T$. 
        \end{assumption}   
    
        Note that if \Autoref{pd} holds, then \Autoref{unique} also holds with high probability, while \Autoref{unique} implies \Autoref{pd}. To show the normality of MLE, we need a regularity assumption below.
        
        \begin{assumption}[Bound on log density] \label{boundeddensity}
            We assume there is a positive constant $c$, s.t. $\min\limits_{i,j \in [p]} (w_i+w_j) > c$. 
        \end{assumption}  
        
        \begin{theorem}[Consistency]
            \label{thm_consistency_mle}
            Under \Autoref{pd}, the MLE satisfies $\widehat{\btheta}_n \to \btheta$ a.s. as $n \to \infty$. As a consequence, $\widehat{\w}_n \to \w$ a.s. as $n \to \infty$.
        \end{theorem}
        
        \begin{theorem} [Asymptotic normality]
        \label{thm_normality_mle}
            Under \Autoref{pd, boundeddensity}, the MLE satisfies $\sqrt{n}(\widehat{\btheta}_n - \btheta) \to N(0, I(\btheta)^{-1})$ in distribution. Hence $\sqrt{n}(\widehat{\w}_n - \w) \to \mathcal{N}(0, BI(\btheta)^{-1}B^T)$ in distribution, where $I(\btheta)$ is the Fisher information matrix. 
        \end{theorem}            
        
        
        Next, we briefly introduce two algorithms to compute the MLE. It can be verified that $l_n(\w)$ is concave, and thus we need to solve the convex optimization problem
        \begin{equation}
            \label{obj}
            \max\limits_{\w \in \mathcal{K}} \ l_n(\w) ,
        \end{equation}
        where $\mathcal{K}$ is the feasible range of $\w$ introduced at the beginning of this subsection. 

        A general algorithm to find an approximation of MLE is the expectation-maximization (EM) algorithm~\cite{dempster1977maximum}, where we treat $X$ as missing. Applying the EM algorithm to \Autoref{loglikelihood}, we can derive the iterative updating formula as 
           $ 
                w_{j}^{(m+1)} = n^{-1}\sum\limits_{i=1}^n\mathbb{P}_m(X_{i}=j \mid  a_i),
            $ 
        where $\mathbb{P}_m(X_{i}=j \mid  a_i) = w_{j}^{(m)} \v_{a_i, j}/( \v_{a_i}^\T\w^{(m)})$ is the conditional probability under $w^{(m)}$ at the $m$-th iteration. A similar update rule was derived for estimating distribution functions from interval-censored data~\cite{wellner1997hybrid}.  
        Another algorithm is the iterative convex minorant (ICM) algorithm developed in survival analysis~\cite{wellner1997hybrid}, which theoretically guarantees the convergence to MLE. In the optimization literature, ICM is also known as the projected Newton-Raphson algorithm~\cite{pan1999extending}.  
            
            
    \subsection{MoM and theoretical properties}
        Calculating the MLE is often computationally expensive, as there is no closed-form solution. Even with the approximation method like EM algorithm, it still needs many iterations to converge. When the design is known, we propose a much faster estimator using the method of moments (MoM) and prove its consistency and asymptotic normality.
        
        We introduce a coefficient matrix $Q=(q_{i j}), q_{i j} = \sum\limits_{i,j \in a}\mu_a, \forall i,j \in [p]$, with the $i$-th row denoted as $q_i$. Note that for the $k$-th coordinate of a random vector $\bm 1_A=(1_{A, 1}, \ldots, 1_{A, p})$, the first order moment 
        $$\gamma_k \de \E ( 1_{A, k} ) = \sum\limits_{j=1}^p \biggl( w_j\sum\limits_{a: k,j \in a}\mu_a \biggr) = q_k^T \w .
        $$ 
        Hence, $\bgamma \de (\gamma_1, \ldots, \gamma_p)^T = Q\w$ establishes the relationship between the first order moment of $\v_A$ and population $\w$.   The empirical first-order moment $\widehat{\bgamma} =n^{-1} \sum\limits_{i=1}^n \bm 1_{a_{i}}$ is the sample mean of $1_{A}$. The above motivates the following moment-based estimator
        \begin{equation}
            \label{gen_mom}
            Q\widehat{\w} = \widehat{\bgamma}.
        \end{equation}        

        \Autoref{pd} ensures that $Q$ is positive definite, so $\widehat{\w} = Q^{-1} \widehat{\bgamma}$.
        We introduce
        $$
        C = Cov(A) = H -\bgamma\bgamma^T, 
        \textrm{ where } H =(h_{i j} ), h_{i i}= \gamma_i, h_{i j}= \sum\limits_{k=1}^p \biggl( w_k\sum\limits_{a: i,j,k \in a}\mu_a \biggr).
        $$ 
        Then, as a direct application of the central limit theorem, we have
        \begin{theorem}
            Under \Autoref{pd}, the MoM estimator defined in \autoref{gen_mom} satisfies $\sqrt{n} (\widehat{\w}-\w) \to N(0, Q^{-1}CQ^{-1})$ in distribution.
        \end{theorem}                
        
        
        In particular, for the uniform design, there is a much simpler form. We have
        $$
        q_{ij} = r_p^{-1}, \quad
        r_p \de \frac{2^{p-1}-p-1}{2^{p-2}-p+1}, \quad \forall i \neq j,
        $$ therefore
        \begin{equation}
            \label{mom}
            \widehat{w}_i+\sum\limits_{j \neq i}\frac{\widehat{w}_j}{r_p} = \hat{\gamma}_i, \quad \text{or equivalently, } 
            \quad \widehat{\w} = \frac{r_p \hat{\bgamma} - \v_{p}}{r_p-1}.
        \end{equation}
        The equivalence is due to $\sum\limits_i \hat{w}_i=1$.

    \subsection{One-step estimator based on MoM}
        Based on the previous MoM estimator, which has a $n^{-1/2}$ rate of convergence, we can further construct an asymptotically efficient estimator (just like the MLE) using the following one-step update.
        \begin{gather}
            \label{one_step}
            \widehat{\btheta}_{one} = \widehat{\btheta}_{MoM}+\nabla^2 l_n(\widehat{\btheta}_{MoM})^{-1}\nabla l_n(\widehat{\btheta}_{MoM}),  \quad
            \widehat{\w}_{one} = \balpha+B\widehat{\btheta}_{one},
        \end{gather}        
        where $\nabla^2$ is the Hessian operator and $\nabla$ is the gradient operator. 
        
         The asymptotic property of the one-step estimator has been well-established for general parameter estimation~\cite[Theorem 20]{ferguson2017course}, so we provide the following result without proof.
        \begin{theorem}
            Under \Autoref{pd, boundeddensity}, the one step estimator defined in \autoref{one_step} has the same asymptotic distribution as MLE.
        \end{theorem}          
        
        
    \subsection{Extension to high-dimensional case}
        \label{subsec:high-dim-mom}
        We show that when $p$ grows with $n$, the MoM estimator is also consistent under some conditions. 
        \begin{theorem}
            \label{thm_mom_dim}
             Let MoM estimator be $\widehat{\w}$. If we use uniform design, then $(p^2\ln p)/n \to 0$ guarantees that $\norm{\widehat{\w}-\w}_1 \to 0$ in probability as $n \to \infty$. Also, with uniform design, $p/n \to 0$ guarantees that $\norm{\widehat{\w}-\w}_2 \to 0$ in probability as $n \to \infty$.
        \end{theorem}        
        
        Also, it can be proved using the Berry-Esseen Theorem \cite{berry1941accuracy} that each coordinate of the MoM estimator is asymptotically normal under the uniform design, even when $p > n$. We omit the technical development here as it is beyond the scope of this paper.   
        
        
\section{Contingency Table for Subset Data}
\label{sec_2D}
        As we mentioned in~\Autoref{subsec_product}, the product mechanism is more suitable for multi-variable cases. In this section, we study the two-variable case by introducing a contingency table for subset data and then proposing methods for independence tests.
        
    \subsection{Formulation}
        Suppose that we have two random variables $X \sim p_{\w_X}, Y \sim p_{\w_Y}$ with $\w_X \in \mathbb{R}^p, \w_Y \in \mathbb{R}^q$, and the joint distribution $\P(X=\x, Y=\y) = w_{\x \y}$, denoted by $(X, Y) \sim p_W$ with $W \in \mathbb{R}^{p\times q}$. By the~\Autoref{prod_design} of product design, if we let both $X \to A$ and $Y \to B$ be conditional designs, then $(X, Y) \to (A, B)$ is a product design. Let $\{ \mu_a^A, \forall a \in \mathcal{A} \}, \{ \mu_b^B, \forall b \in \mathcal{B} \}$ denote two conditional designs and $s=\abs{\mathcal{A}}, r=\abs{\mathcal{B}}$ denote the cardinalities of possible outcomes. A typical $s$ is much larger than $p$, and it is at most $2^p-p-2$.
        
        The observations under this setting are $n$ i.i.d. pairs of $A_i, B_i$. Thus, it is natural to consider a two-way contingency table for $A$ and $B$. In particular, the table $N = (\{n_{a b}\})_{s \times r}$, where each $n_{a b}$ is the number of observations satisfying $A=a, B=b$. 

    \subsection{Independence test}

        We study how to perform independence testing between $X$ and $Y$ based on the subset-valued contingency table $N$. 
        Our null hypothesis is $H_0: X \indep Y$ against the alternative $H_1: X \nindep Y$. 
        It can be verified that the joint and marginal distributions for $A, B$ are
        \begin{gather*}
            \P(A=a, B=b) = \mu_a^A\mu_b^B \v_a^TW\v_b,\quad
            P(A=a)=\mu_a^A \v_a^T\w_X, \quad
            P(B=b)=\mu_b^B \v_b^T\w_Y .
        \end{gather*}
        Hence, the above hypothesis is equivalent to $H_0: A \indep B$ against $H_1: A \nindep B$. Based on this equivalence, we develop four tests below and numerically compare them in \Autoref{sec_exp}. 
        

        The methods are based on the estimation of distributions. With the contingency table, we can consistently estimate the joint distribution $W$ using any of the methods mentioned in \Autoref{sec_est} with a subtle modification. 
        Let $\widehat{W}, \widehat{\w}_X, \widehat{\w}_Y$ be the estimated parameters from \Autoref{sec_est}. We then estimate the distributions of $A, B$ with
        $$
        \widehat{\P}(A=a, B=b) = \mu_a^A\mu_b^B \v_a^T\widehat{W}\v_b, \quad
        \widehat{\P}(A=a)=\mu_a^A \v_a^T\widehat{\w}_X, \quad \widehat{\P}(B=b)=\mu_b^B \v_b^T\widehat{\w}_Y .
        $$
        We note that some general independence testing methods developed for continuous random variables~(e.g., \cite{szekely2007measuring,wang2017generalized,zhang2019bet}) may also be adapted to handle our subset data, and we leave it for future work. 
    
        \subsubsection{Pearson's Chi-square test}
            The classical Pearson's Chi-square test statistics in our context is 
            $ 
                T_{P} = \sum_{a, b} (n_{a b}-e_{a b})^2/e_{a b} ,
            $ 
            where $e_{a b} = n \widehat{\P}(A=a)\widehat{\P}(B=b)$ is the expected number of observations under the null hypothesis.
            Under the null hypothesis, $T_P$ asymptotically converges to $\chi^2_{s r}$ as $n \to \infty$. A common rule (and also our experiments) suggest using Pearson's Chi-square test only when all $n_{a b} \geq 5$. 

        \subsubsection{Likelihood ratio test (LRT)}
            Recall that the joint distribution of $A, B$ is decided by $W$. The  LRT here is based on the estimation of $W$ and $\w_X, \w_Y$. In other words, the null is $H_0: W = \w_X \odot \w_Y$ against $H_1: W$ represents a free discrete distribution. The test statistic is 
            \begin{equation}
                T_{L} = 2\sum_{a, b} n_{a b} \ln \frac{\v_a^T\widehat{W}\v_b}{\v_a^T\widehat{\w}_X \widehat{\w}_Y^T\v_b} . \label{eq_LRT}
            \end{equation}
            
            Under the null hypothesis, $T_L$ converges to $\chi^2_{(p-1)(q-1)}$ in distribution as $n \to \infty$. Hence, the limiting distribution has a smaller degree of freedom compared with $T_P$.
        
        \subsubsection{Approximated LRT}
            From the experimental results of parameter estimation methods in \Autoref{sec_exp}, we find that the MoM estimator is significantly faster than other estimators but with similar accuracy. This motivates us to replace the MLE with MoM when evaluating the LRT. 
            This leads to a test statistic the same as \Autoref{eq_LRT}, except that the distribution parameters are estimated by MoM. 
            Though the above test is to approximate the LRT, we experimentally find that the statistical power does not degrade. 
        
        \subsubsection{Bonferroni correction}
            Instead of testing $H_0: A \indep B$, we simultaneously test the independence for each pair of coordinates of $\v_A = (\v_{A,1}, \ldots, \v_{A,p})^T, \v_B = (\v_{B,1}, \ldots, \v_{B,q})^T$ with a Bonferroni correction. In other words, we test $H_0: \v_{A,x} \indep \v_{B,y}, \forall x=1,\ldots, p, y=1,\ldots,q. $
            It proceeds as follows. For each $x=1,\ldots, p, y=1,\ldots,q$, we use the Pearson's Chi-square to calculate a \textit{p}-value $p_{xy}$ for $H_0: \v_{A,x} \indep \v_{B,y}$. The corresponding aggregated contingency table has the form in \Autoref{tab:bonferroni_table}.
            With a significance level $\alpha$, we reject the null hypothesis if $\min\limits_{1\leq x \leq p, 1\leq y \leq q} p_{x y} < \alpha/pq$.

            \begin{table}[tb]
                \centering
                \caption{Aggregated Contingency Table}
                \label{tab:bonferroni_table}
                \begin{tabular}{|l|c|c|}
                    \hline
                    \diagbox[]{$\v_A^{(x )}$}{$\v_B^{(y)}$}& 1& 0 \\ \hline
                    1& $\sum\limits_{a,b:x \in a, y\in b} n_{a b}$ & $\sum\limits_{a,b:x \in a, y\notin b} n_{a b}$\\ \hline
                    0& $\sum\limits_{a,b:x \notin a, y\in b} n_{a b}$& $\sum\limits_{a,b:x \notin a, y\notin b} n_{a b}$\\ \hline
                \end{tabular}
            \end{table}

\section{Experimental Study on the Adult Dataset}
\label{sec_exp}
        In this section, we provide a real-data example to demonstrate the subset privacy and developed methods. More experimental results on simulated data are provided in the Appendix~\ref{sec_exp_simu}. 
        
        We use the `Adult' (also known as `Census Income') dataset, which Barry Becker extracted from the 1994 Census database \cite{Dua:2019}. The dataset contains $N=32561$ observations with $14$ attributes. Among those attributes, race, gender, and income are of particular interest. 
            
        \subsection{Distribution of race}
            There are five categories of race. The numerical labels and population distribution $\w$ are summarized in \Autoref{tab:race_pop_dist}. Sometimes, race is considered as sensitive information, and it can be protected by subset privacy. Here, we show that our method provides desirable privacy, and it is efficient to estimate the population distribution.
            
\begin{figure}[tb]
                \begin{minipage}{0.45\textwidth}
                    \centering                        \includegraphics[width=\textwidth]{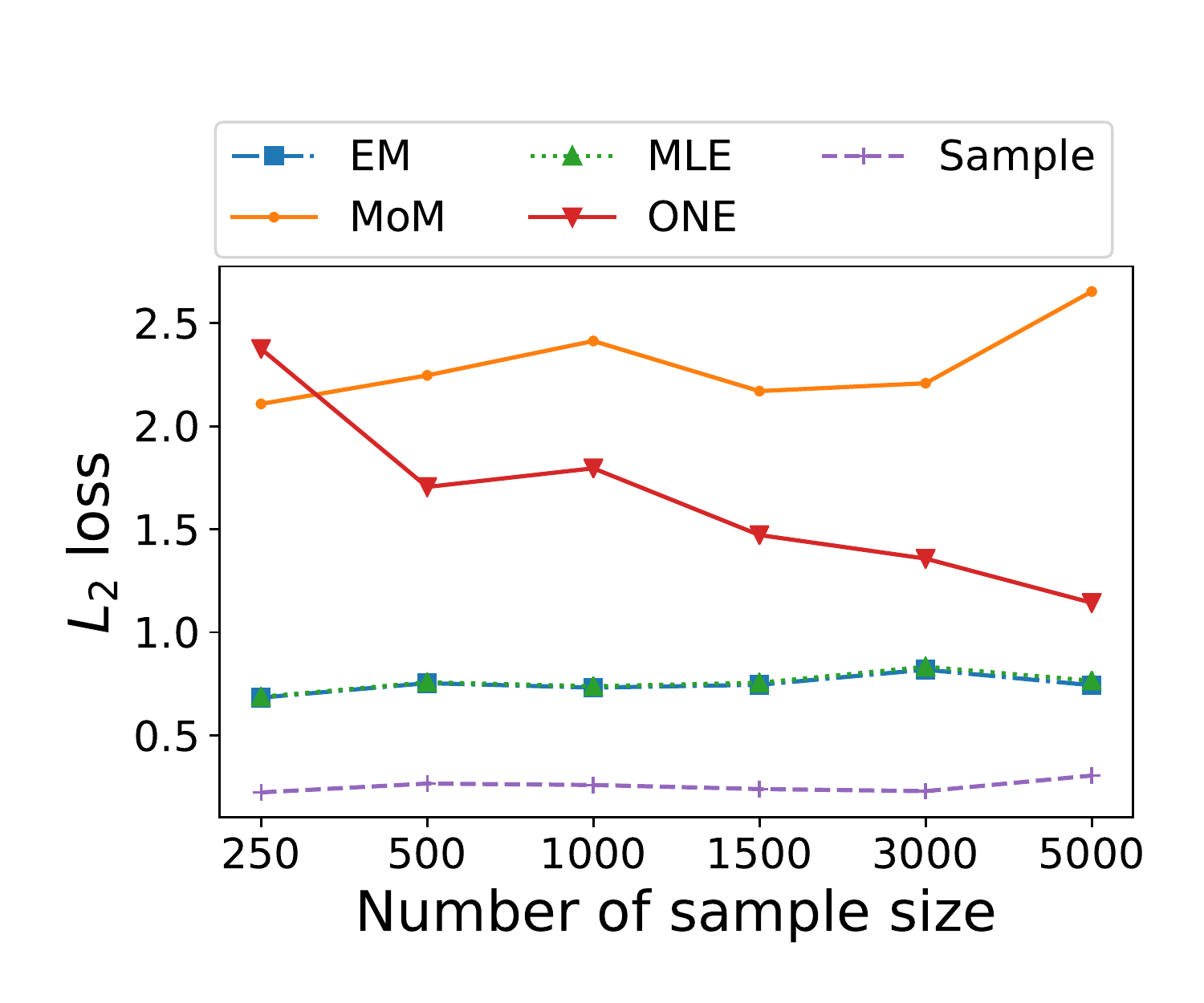}
                    \vspace{-0.4in}
                        \caption{Mean scaled $L_2$ loss for different estimators as $n$ increases, for $100$ replications.}
                        \label{fig:adult_race_l2_loss}
                \end{minipage}
                \hfill
                \begin{minipage}{0.45\textwidth}
                    \centering
                        \includegraphics[width=\textwidth]{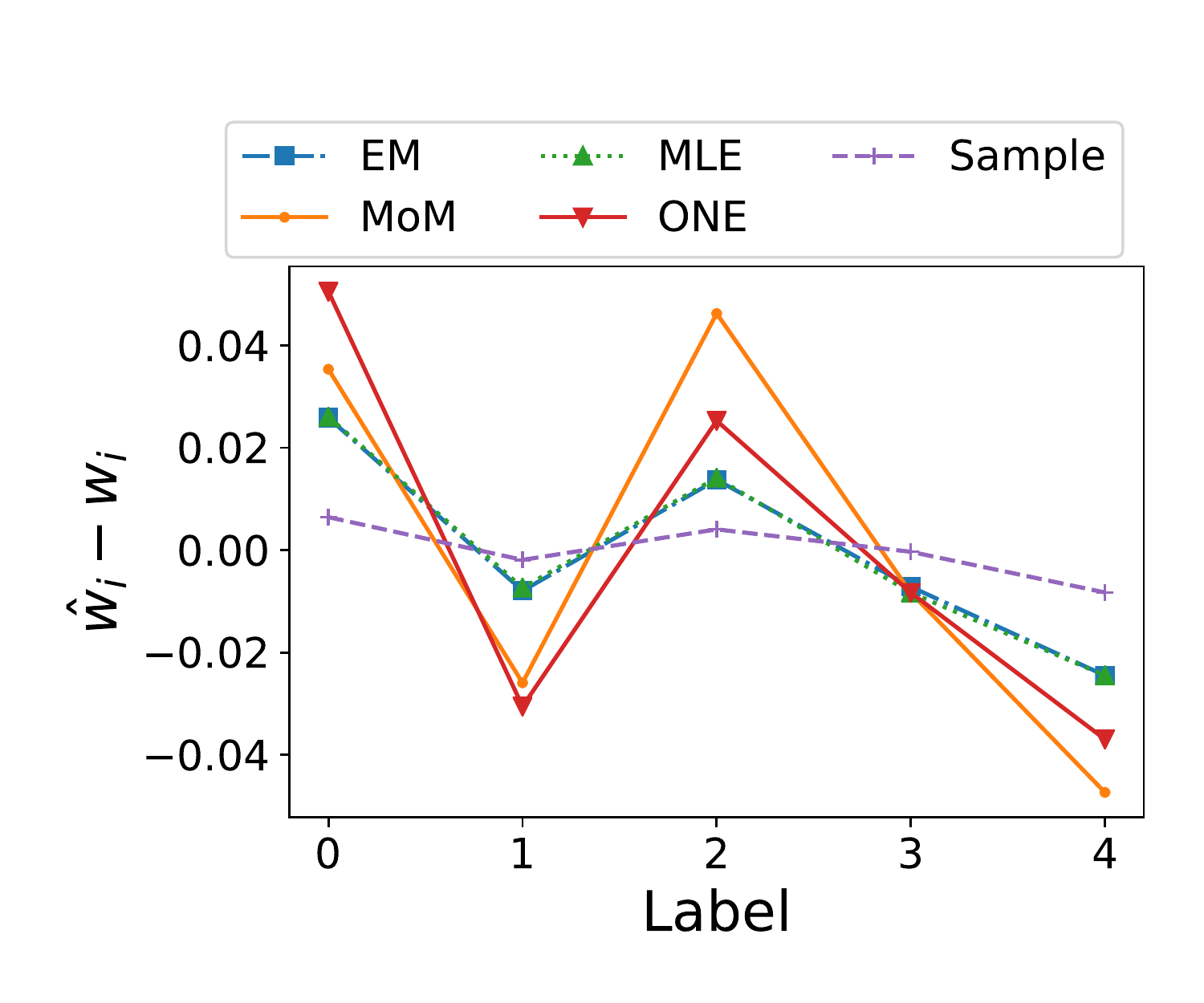}
                        \vspace{-0.4in}
                        \caption{The difference of estimated and population distribution for one run, with $n=500$.}
                        \label{fig:race_diff}
                \end{minipage}
            \end{figure}

            In the experiments, we randomly draw $n$ samples from the original dataset in each replication. We draw subset observations based on the uniform design and estimate $\hat{\w}$. To stabilize the estimation, we replicate $k=100$ times and record the average scaled $L_2$ loss 
            $n\norm{\widehat{\w}-\w}_2^2$. 
            Its expectation is expected to converge to a constant as $n\to\infty$ for the proposed methods. We implement and compare the estimators in \Autoref{sec_est}, including the MLE estimated by the EM algorithm (`EM'), the MLE solved by a general-purpose optimization package \textit{CVXPY}~\cite{diamond2016cvxpy} (`MLE'), the moment-based estimator (`MoM'), and the one-step estimator derived from MoM (`ONE'). The additional line named `Sample' is calculated from the original non-private data. The results for different values of $n$ are summarized in \Autoref{fig:adult_race_l2_loss}. We also visualize the difference between estimates and `true' parameters calculated from the whole dataset (in hindsight) in one replication (\Autoref{fig:race_diff}). 
            
            From the results in \Autoref{fig:adult_race_l2_loss}, the $n^{-1/2}$ rate of convergence holds even for a relatively small sample size. 
            Also, the $L_2$ loss of MLE is about four times the loss of `Sample', indicating that we need to double the sample size to obtain an error rate comparable with that of non-privatized data. 
            Next, we will show the amount of privacy in the privacy mechanism. 
            
            \begin{table}[tb]
            \caption{Population Distribution and Labels of Race}
                \label{tab:race_pop_dist}
                \centering
                \vspace{0.1in}
            \scalebox{0.9}{
            \begin{tabular}{lll}
            \toprule
            \textbf{Numerical label} &                 \textbf{Race category} &         \textbf{$\w$} \\
            \midrule
            0 &   Amer-Indian-Eskimo &  0.009551 \\
            1 &   Asian-Pac-Islander &  0.031909 \\
            2 &                Black &  0.095943 \\
            3 &                Other &  0.008323 \\
            4 &                White &  0.854274 \\
            \bottomrule
            \end{tabular}
            }
            \end{table}

        \subsection{Amount of privacy}
\begin{figure}[tb]
                    \centering \includegraphics[width=0.65\textwidth]{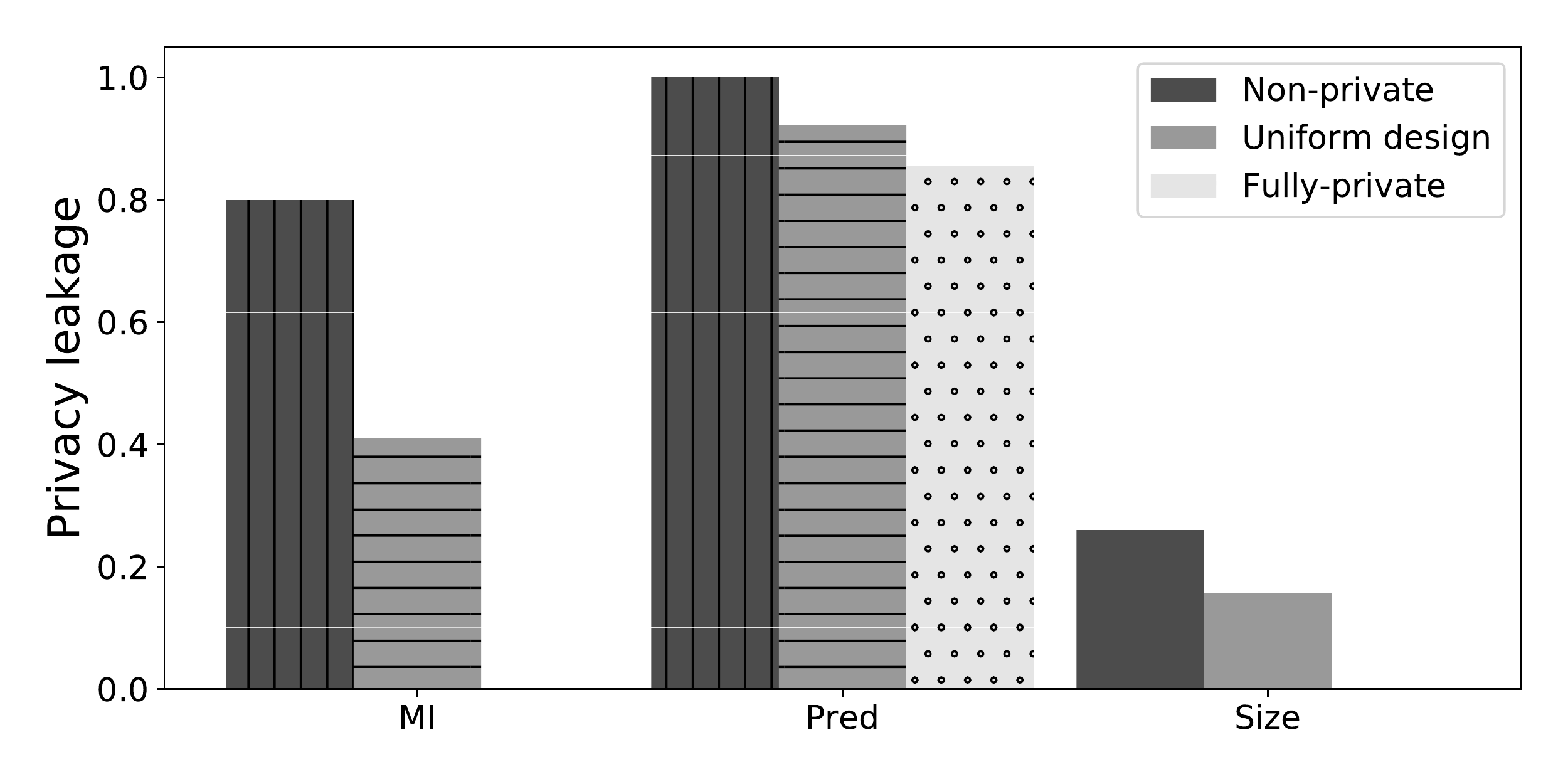}
                    \vspace{-0.2in}
                    \caption{Comparison of the privacy leakage for three designs, as evaluated by mutual information (`MI'), prediction risk (`Pred'), and subset size (`Size').
                    Note that the MI and Size for the fully-private are zero in the plot.
                    } 
         \label{fig:adult_privacy_leak}
            \end{figure}

            We calculate the privacy leakage for three designs using the three ways introduced in \Autoref{subsec_quantity}. The three designs include the non-private design ($A=\{X\}$), uniform design, and fully-private design ($A=\textrm{range}(X)$). The results are shown in \Autoref{fig:adult_privacy_leak}.
            
            As we expect, the privacy leakage of the uniform design is within the range of no privacy and full privacy. Mutual information privacy leakage is half the maximum leakage, meaning that the information about $X$ contained in subset observations is one bit (either in or not in the subset). Size privacy leakage decreases from $0.25$ to $0.15$ after the use of uniform design. The interpretation is that we can eliminate $0.25$ probability by observing the non-private data $X$ on average, while we can only eliminate $0.15$ from the private data $A$. Prediction privacy leakage decreases from $1$ to $0.9$, meaning that the probability we correctly guess the label is $0.9$ from the uniform design. Without any privatization, we immediately know the true label, so the leakage is $1$. 
            One may worries that $0.9$ represents a considerable leakage. However, we note that it needs to be compared with the minimum leakage. 
            If we do not know any knowledge about $X$, then the optimal strategy is guessing the most likely category, which is `white' in this case, with the average accuracy as $0.85$. 
            
            In the above results, the measurement of privacy is averaged over the population. Recall that the size and prediction privacy can be applied to evaluate each individual's privacy as well. We illustrate the idea below. 
            
\begin{figure}[tb]
                    \centering \includegraphics[width=\textwidth]{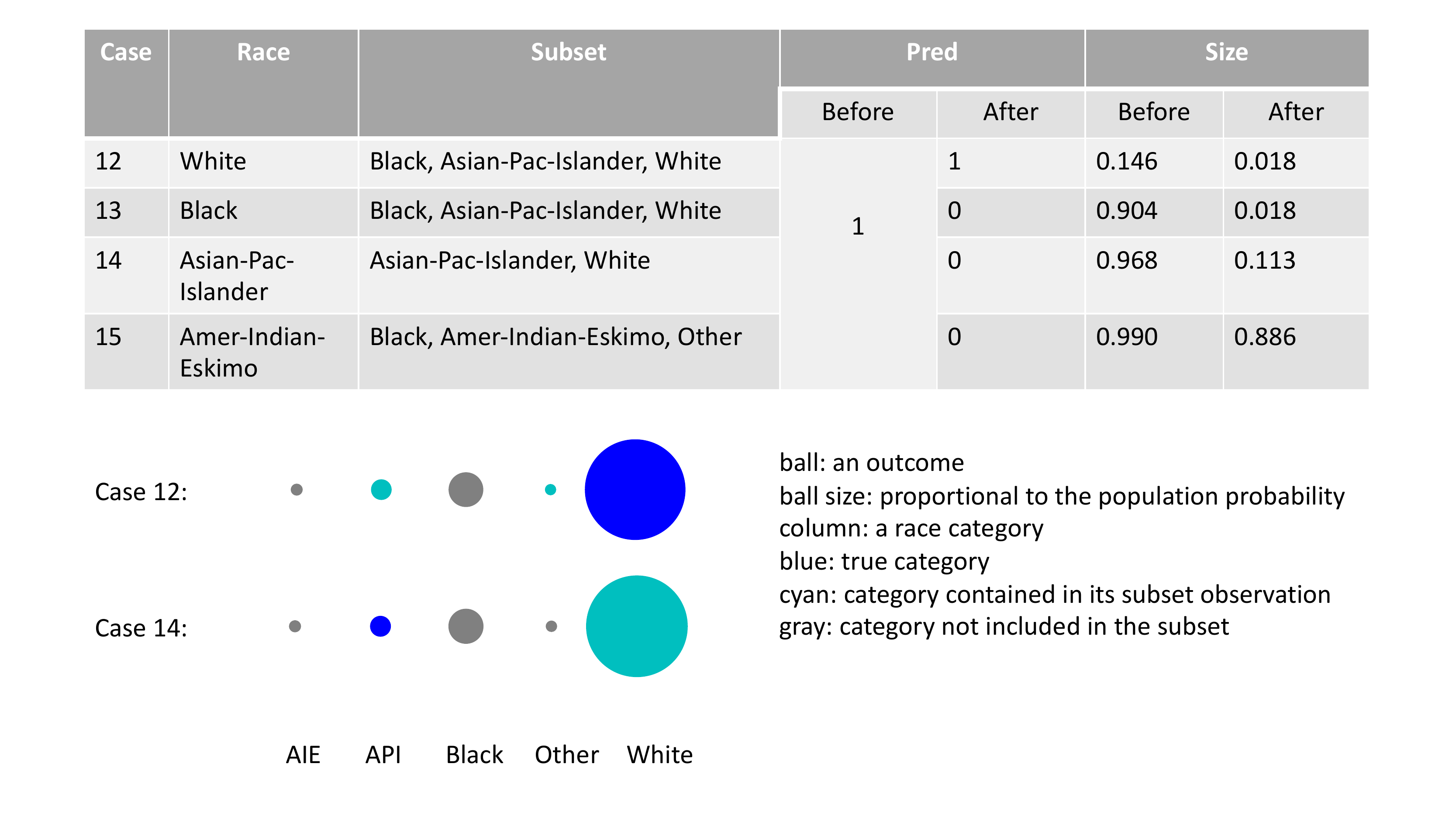}
                    \vspace{-0.5in}
                    \caption{Privacy leakage at the individual level for a snapshot of the dataset (top) and graphical illustrations for size privacy leakage (bottom). The evaluation is based on either prediction privacy leakage (`Pred') or subset size leakage (`Size'). } \label{fig:adult_privacy}
            \end{figure}

            \Autoref{fig:adult_privacy} shows a realization of the subset-private data of four observations from Case 12 to 15. We examine Case 12 and 14 in particular. The race of Case 12 is white, and the observed subset consists of black, white, and Asian. The optimal prediction is the category with the largest estimated probability (white, $0.854$). Because the true category coincides with the most likely one, the prediction leakage is one despite the privatization. 
            The above scenario will change for Case 14, whose category is Asian-Pac-Islander. The subset observation includes white, 
            so the optimal guess is still white, and an adversary will not coincidentally recover the true label. Moreover, a large amount of uncertainty is brought in because of the inclusion of white in the subset. 
            The subset size leakage decreases significantly from 0.97 to 0.11 (compare the ball size of gray plus cyan and gray solely). 
            
            In conclusion, subset privacy protects the privacy of rare categories much better than the frequent categories at the individual level, which is an attractive property in real-world applications.
            
        \subsection{Independence tests of gender and income}\label{sec_gender_income}
\begin{figure}[tb]
                    \centering
            \begin{minipage}[b]{0.5\textwidth}
                    \centering          
            \captionsetup{type=table}
                \caption{Contingency Table of Gender and Income for the Whole Dataset}
                \vspace{0.1in}
                \label{tab:cont_full}
                \centering
                \scalebox{0.8}{
                \begin{tabular}{lrr}
                \toprule
                Gender & Income=`Low' & Income=`High' \\
                \midrule
                Female   &   9592 &  1179 \\
                Male   &  15128 &  6662 \\
                \bottomrule
                \end{tabular}
            }
            \end{minipage}
            \hfil
            \begin{minipage}{0.4\textwidth}
                    \centering          
             \includegraphics[width=\textwidth]{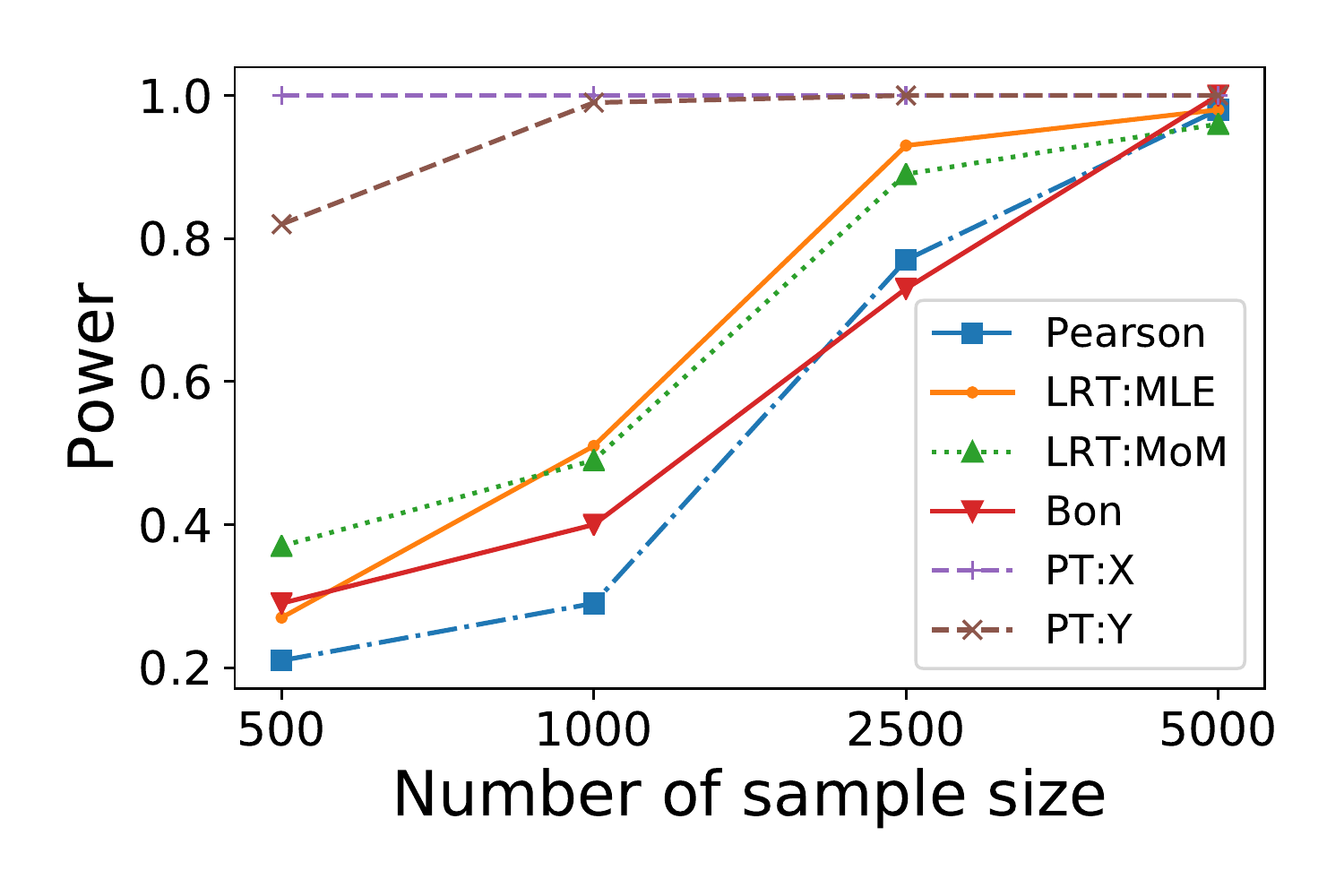}
                    \vspace{-0.4in}
                    \caption{The statistical power of four independence tests (null: gender is independent with income), given the significance level $\alpha=0.05$, $k=100$, and varying $n$.}
                 \label{fig:adult_power}
            \end{minipage}
            \end{figure}

            The last application is to test the independence of gender and income under subset privacy. The null hypothesis is that they are independent. The contingency table from the whole non-private dataset is presented in \Autoref{tab:cont_full}. Pearson's Chi-square test has a \textit{p}-value close to zero, indicating that the two variables are effectively dependent at the population level. In the experiment, we take the dependence between gender and income as the ground truth. 
            
            We will show the power of our tests while controlling the probability of type I error to be $5\%$. 
            Specifically, we consider the Pearson's Chi-square test (`Pearson'), likelihood ratio test designed for subset privacy (`LRT:MLE'), likelihood ratio test with the MLE approximated by MoM (`LRT:MoM'), Bonferroni correction-based test (`Bon').
            The subset data are generated as follows. First, since both gender and income are binary variables, after sampling $n$ observations from the original dataset, we apply `$p=2$'-uniform design introduced in \Autoref{subsec_p23} to sample subsets from the Markov chain $X \to Y \to A$. Then, we perform all four tests based on subset observations. For comparison purposes, we also apply Pearson's Chi-square test to the non-private samples $X$ and intermediate data $Y$, denoted by `PT:X' and `PT:Y', respectively. 
            We independently replicate the above procedure $k$ times, and each time we calculate the \textit{p}-value and power for each test. The results are summarized in \Autoref{fig:adult_power}. More extensive experiments and implementation details are included in the Appendix~\ref{sec_exp_simu}. 
            
            We note that the transformation from the original data $X$ to intermediate data $Y$ loses some information, and the subset sampling further loses information. \Autoref{fig:adult_power} also implies that the sample size of subset observations needs to be four times that of data $Y$ to achieve the same power. When the sample size is sufficiently large, all the methods can identify the dependence between gender and income.
            Overall, the LRT:MLE and LRT:MoM are better than other tests.
            The above results are consistent with our experimental results on simulated data (in the Appendix~\ref{sec_exp_simu}).
            
\section{Conclusion and Further Remarks}
\label{sec_con}
    We proposed a new local data privacy methodology for collecting categorical data. The main idea is to obfuscate the original category with a random subset. 
    We studied the following aspects of subset privacy. 
    First, for parameter estimation, we proposed a computationally efficient method that attains the same asymptotic efficiency as MLE. Theoretical analysis is provided to evaluate the identifiability, estimability, and asymptotic performance.
    Second, we developed metrics to quantify privacy leakage. 
    Finally, we studied subset data-based independence testing by adapting some classical methods. 
    
    Some interesting future topics related to subset privacy include the following. First, we studied the use of subset-private data for parameter estimation and testing. It remains a challenge to perform other learning tasks such as regression and classification involving such a data type. 
    Second, for high-dimensional data, the computation cost will be heavy under the proposed independent mechanism. It is important to develop new mechanisms with fewer subset candidates. 
    Also, it would be interesting to study the optimal utility-privacy tradeoffs under subset privacy.

\appendices


\section*{Appendix}

\section{Further Discussions on the Comparison with Differential Privacy}
In the following, we highlight some subtle aspects of subset privacy and elaborate on its connections with (local) differential privacy. 

\begin{enumerate}
    \item \textit{The main advantages of subset privacy}. 
The subset privacy offers the following practical benefits. First, it is easy to implement since its data collection procedure is naturally compatible with the existing survey-based collection methods. Second, it is user-friendly as individuals will only need to select from randomly-generated subset choices (e.g., through an open-source interface). Third, the collected message will obfuscate but not distort the original message, which can be extremely helpful in application domains that require authentic information (e.g., in the census). The above features distinguish subset privacy with (local) differential privacy and its variants. 
    
    \item \textit{Rigorous privacy guarantee}. We will extend the aforementioned subset mechanisms to actively control size privacy at an individual level in Appendix~\ref{sec:dummy}. It allows us to have a user-chosen lower bound of the size privacy regardless of the raw data. 
    On the other hand, the subset privacy can be implemented in a population-independent manner, as the independent mechanism does not rely on the knowledge of the underlying data distribution.
    As a result, both the implementation and the coverage guarantee of subset privacy do NOT depend on the population distribution. This favorable property is also shared by differential privacy.
    
    \item \textit{The threat of side information}. The advantage of keeping the truth comes with the cost. Subset privacy is not robust to side information. We need to assume that the adversary does not know the identity of any subset before collecting data (\Autoref{subsec_quantity}). Nevertheless, our threat model still covers a wide range of application situations where data are collected from anonymous individuals. 
\end{enumerate}

\section{Control of Privacy Coverage}
\label{sec:dummy}
    In this section, we propose a modification of subset privacy, which enables us to control the privacy coverage. In other words, we can set an arbitrary lower bound (within $[0,1/2)$) for the size privacy coverage. 
    
    We consider one-dimensional variable $X \in \mathbb{R}^p$ for brevity. The technique can be extended to the multi-dimensional case using the product design. We further consider the independent design $\{ \nu_a,  a \in \mathcal{A} \}$ with $\nu_a = \nu_{a^c}$. For example, we may use a uniform design. The main idea is enlarging the domain of $X$ from $[p]$ to $[p+2]$, while the last two categories are dummy or artificial categories. Then, we can generate dummy users with dummy categories. We also modify the independent design to ensure that the subset always contains one of the dummy categories. In this way, given a subset, an adversary can never tell whether it comes from a dummy or true user. 
    We can control the ratio of dummy categories and thus the privacy coverage. The technical details are described below.  
    
    For a parameter $0<\alpha<1/2$ that gives the lower bound of the privacy coverage, we normalize the population distribution by $\hat{w}_j = (1-2\alpha)w_j, j \in [p]$, and $w_j=\alpha, j=p+1,p+2$. Since the probability of a dummy category is at least $\alpha$, and a subset must contain one of them, the size coverage is lower bounded by $\alpha$. 
    Accordingly, we state the following new subset-generating mechanism. If $X$ is a true user ($X \in [p]$), then we draw a subset $A \subset [p]$, $X \in A$ by using the original design $\{ \nu_a,  a \in \mathcal{A} \}$. Then, we add one of the dummy categories to $A$ with equal probability to produce the final output subset $\hat{A}$. If $X$ is dummy, we draw $\tilde{A}$ from $\{ \nu_a,  a \in \mathcal{A} \}$, and $\hat{A}=\tilde{A} \cup X$. It can be seen that this procedure defines a new conditional design on the enlarged data domain $[p+2]$. Hence, we can apply all the developed methods to analyze the enlarged subset $\hat{A}$.
    
    In summary, in addition to collecting the data from $n$ users, we also simulate $2\alpha n$ dummy users, each associated with a dummy category. Then, the collection system creates subsets from the induced design stated above. As long as $\alpha$ is revealed, the inference of true users' population information can be derived.  Such modification ensures minimum size privacy to be $\alpha$, retaining the virtue of faithfulness. Users only have to answer a relatively non-sensitive question of whether the true value is contained in a subset with a guaranteed privacy coverage. 

\section{Two- or three-category Variables}
\label{sec_p23_alt}
    Using a similar technique proposed in \Autoref{sec:dummy}, we introduce another way to address two- or three-category variables. We only have to allow the independent design to sample subsets $A$ such that $\abs{A}=1$ or $\abs{A}=p-1$. For instance, when $p=2$, if $X \in [p]$, then $\hat{A}=\{X, 3\}$ or $\hat{A}=\{X, 4\}$ with equal probability; if $X \notin [p]$, then $\hat{A}=\{X, 1\}$ or $\hat{A}=\{X, 2\}$ with equal probability.

\section{More Experiments on Simulated Data}
    \label{sec_exp_simu}
    In this section, we evaluate the developed methods and computation costs using simulated data. 
    The experiments are based on independent mechanisms, and we use $\{ \mu_a, a \in \mathcal{A} \}$ to denote the conditional design induced by an independent design $\{ \nu_a,  a \in \mathcal{A} \}$. 
          
    \subsection{Performance of estimators} \label{subsec_est_exp}
        We implement and compare the estimators in \Autoref{sec_est}, including the MLE estimated by the EM algorithm (`EM'), the MLE solved by a general-purpose optimization package \textit{CVXPY}~\cite{diamond2016cvxpy} (`MLE'), 
        the moment-based estimator (`MoM'), and the one-step estimator derived from MoM (`ONE'). 
        In particular, we experimentally show that when $p$ is fixed, these estimators are consistent and asymptotically normal with the derived asymptotic variance.
        We also provide a comparison of the executing time (based on an Intel Core i5 2.3GHz Quad-core CPU, measured in seconds). 
        
        \subsubsection{Consistency and normality}
            The data is generated as follows. We simulate for each combination of $p=4, 8$ and $n=100, 500, 1000$, and replicate $k=1000$ times.  For each replication, we choose a uniformly generated population distribution $\w$ and generate the data $X$. We use a uniformly generated independent design $\{\nu_a, a \in \mathcal{A}\}$ to realize the Markov chain $(X, \tilde{A}) \to A$.  
            
            The estimation is evaluated by the scaled $L_2$ loss $n\norm{\widehat{\w}-\w}_2^2$. 
            Its expectation is expected to converge to a constant as $n\to\infty$ for the above four methods.
            In particular, we expect that $n\norm{\widehat{\w}-\w}_2^2 \to \sum\limits_{j=1}^p \lambda_i Z_i , Z_i \overset{i.i.d}{\sim} \chi^2_1$ with $\lambda_i$'s the eigenvalues of $BI(\btheta)^{-1}B^T$, if $\sqrt{n}(\widehat{\w}_n - \w) \to N(0, BI(\btheta)^{-1}B^T)$ holds (for MLE under regularity conditions). Note that the $\lambda_i$'s are decided by $\{\nu_a, a \in \mathcal{A}\}$ and $\w$. Similarly, for the MoM, we expect $\sqrt{n} (\widehat{\w}^{(n)}-\w) \to N(0, Q^{-1}CQ^{-1})$ and $n\norm{\widehat{\w}-\w}_2^2 \to \sum\limits_{j=1}^p \lambda_i Z_i$, where $\lambda_i$'s are the eigenvalues of $Q^{-1}CQ^{-1}$. 
            
            We report the mean loss and its standard error for each combination of $p, n$ in \Autoref{tab:consistency}. MLE:LIM and MoM:LIM represents the theoretical limits of scaled $L_2$ loss for MLE and MoM, respectively. The empirical results of the estimators are aligned with our theoretical derivation. The MoM has a larger asymptotic variance than other estimators. The EM, MLE, and ONE exhibit similarly in this experiment.

            
            \begin{table}[tb]
                \caption{Scaled $L_2$ loss $n\norm{\widehat{\w}-\w}_2^2$ for four estimators, with $p$ and $n$ vary. The experiment is repeated 1000 times and mean loss(standard deviation) is reported. MLE:LIM and MoM:LIM is the theoretical limits of the scaled loss for MLE and MoM.}
                \label{tab:consistency}
                \vspace{0.1in}
            \centering
            \scalebox{0.9}{
           \begin{tabular}{@{}lllllll@{}}
            \toprule
           \multirow{2}{*}{\textbf{Methods}}         & \multicolumn{3}{c}{\textbf{$p=4$}}                                                          & \multicolumn{3}{c}{\textbf{$p=8$}}                                                          \\
            \cmidrule(lr){2-4} \cmidrule(lr){5-7} 
            {}         & \multicolumn{1}{c}{\textbf{$n=100$}} & \multicolumn{1}{c}{\textbf{$n=500$}} & \multicolumn{1}{c}{\textbf{$n=1000$}} & \multicolumn{1}{c}{\textbf{$n=100$}} & \multicolumn{1}{c}{\textbf{$n=500$}} & \multicolumn{1}{c}{\textbf{$n=1000$}} \\ \midrule
            EM        & 2.24(0.07) & 2.51(0.09) & 2.35(0.09) &  5.04(0.09) & 5.57(0.1)  & 5.57(0.1)  \\ 
            MLE       & 2.24(0.07) & 2.51(0.09) & 2.35(0.09) &  5.05(0.09) & 5.58(0.1)  & 5.59(0.1)  \\ 
            ONE & 2.52(0.1)  & 2.61(0.1)  & 2.38(0.09) &  4.99(0.09) & 5.63(0.1)  & 5.64(0.1)  \\ 
            MLE:LIM  & 2.5(0.09)  & 2.56(0.09) & 2.5(0.09)  &  5.68(0.1)  & 5.67(0.1)  & 5.68(0.1)  \\ 
            MoM       & 2.79(0.16) & 2.88(0.13) & 2.66(0.1)  &  6.81(0.12) & 6.78(0.12) & 6.68(0.11) \\ 
            MoM:LIM  & 2.71(0.09) & 2.78(0.1)  & 2.71(0.09) &  6.74(0.11) & 6.74(0.11) & 6.74(0.11) \\ \bottomrule
            \end{tabular}
            }
            \end{table}        
        
\begin{figure}
                \begin{minipage}{0.45\textwidth}
                    \centering
                        \includegraphics[width=0.9\textwidth]{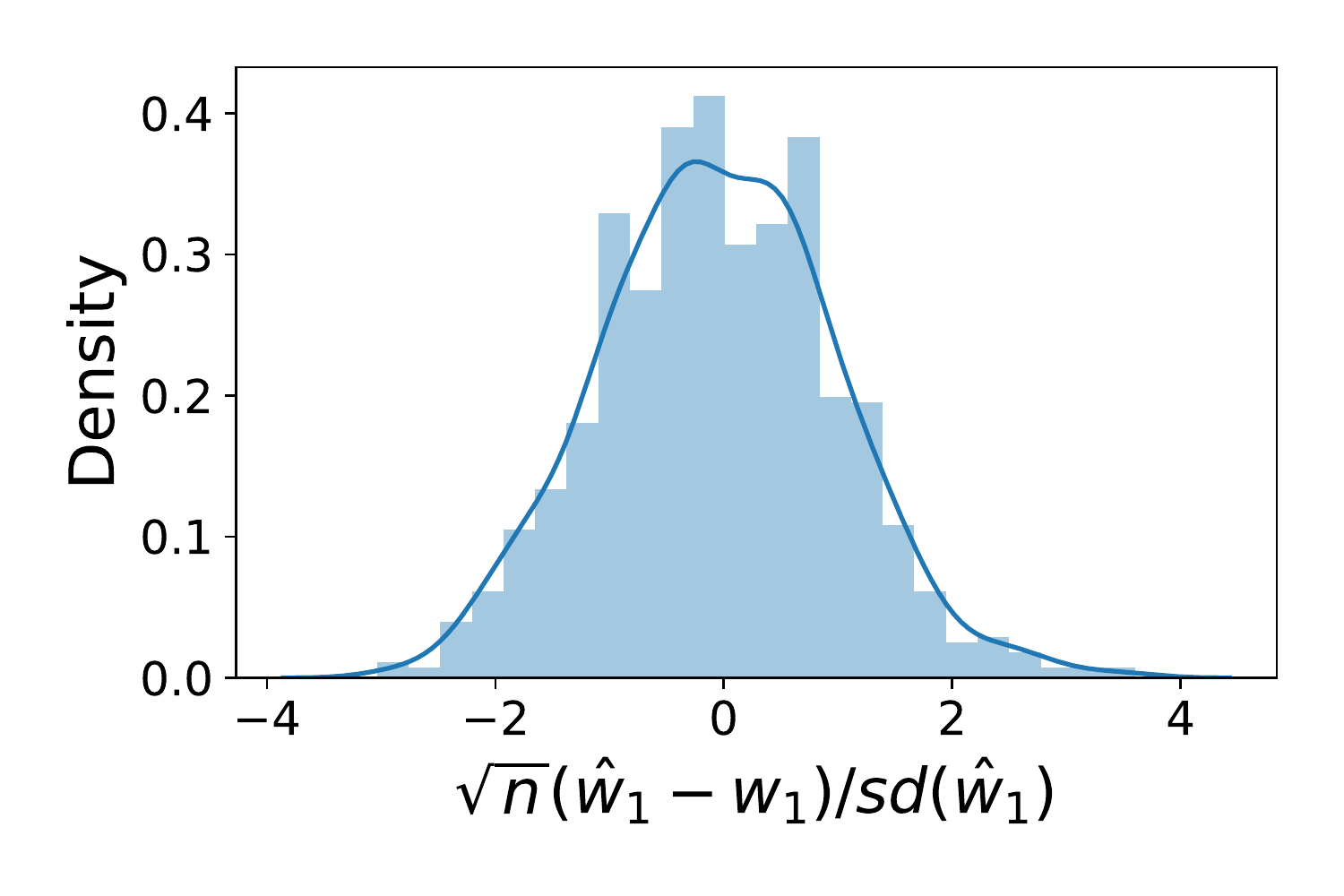}
                \end{minipage}
                \hfill
                \begin{minipage}{0.45\textwidth}
                    \centering
                        \includegraphics[width=0.9\textwidth]{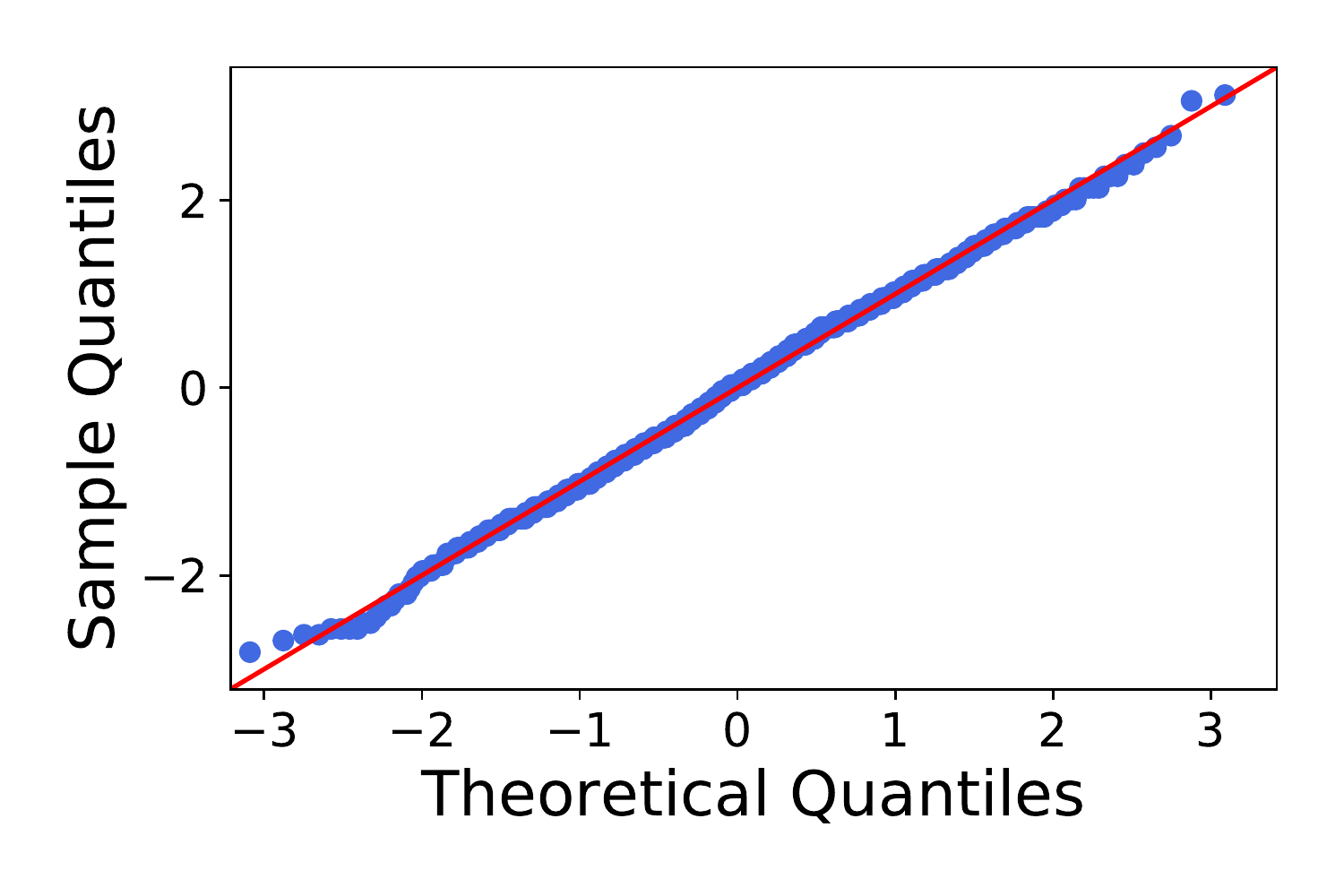}
                \end{minipage}
                \caption{The left plot is the histogram of the empirical distribution of the normalized $\hat{w}_1-w_1$ from the MoM estimator (1000 replications under the uniform distribution and design). 
                The right plot is the corresponding Q-Q plot.}
                \label{fig:dist}
            \end{figure}

            To visualize the normality, we present a typical distribution of $\widehat{w}_i$ from MoM. Similar results can be obtained from other estimators. We use a uniform population distribution and uniform design, with $p=4, n=1000$, and $k=1000$. The histogram and Q-Q plot are displayed in \Autoref{fig:dist}.
            
        \subsubsection{Extension to the high dimension}
\begin{figure}
                \begin{minipage}{0.45\textwidth}
                    \centering
                      \includegraphics[width=0.9\textwidth]{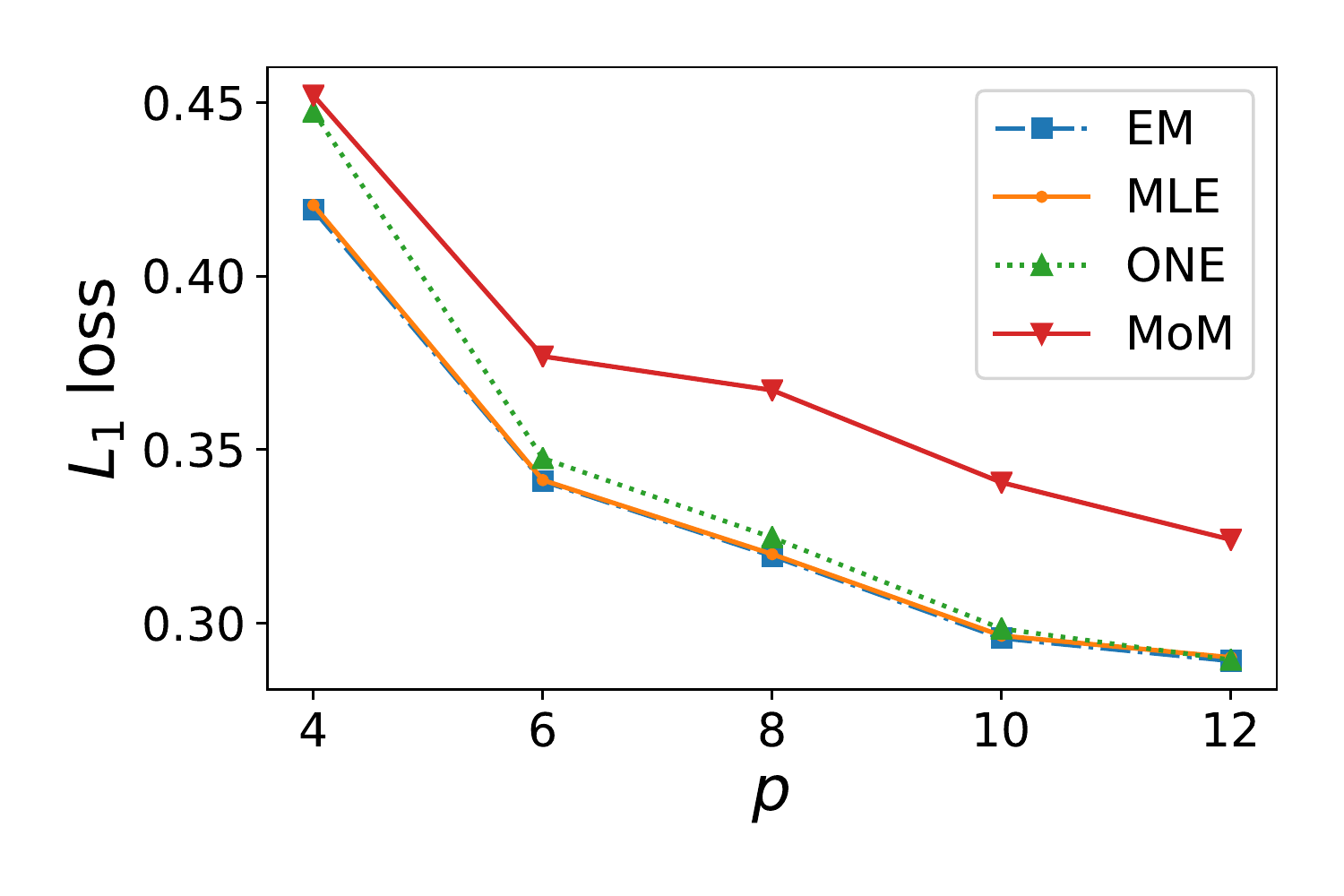}
                        \caption{Average $L_1$ loss for 1000 replications when $p$ and $n$ vary.}
                        \label{fig:high_1}
                \end{minipage}
                \hfill
                \begin{minipage}{0.45\textwidth}
                    \centering \includegraphics[width=0.9\textwidth]{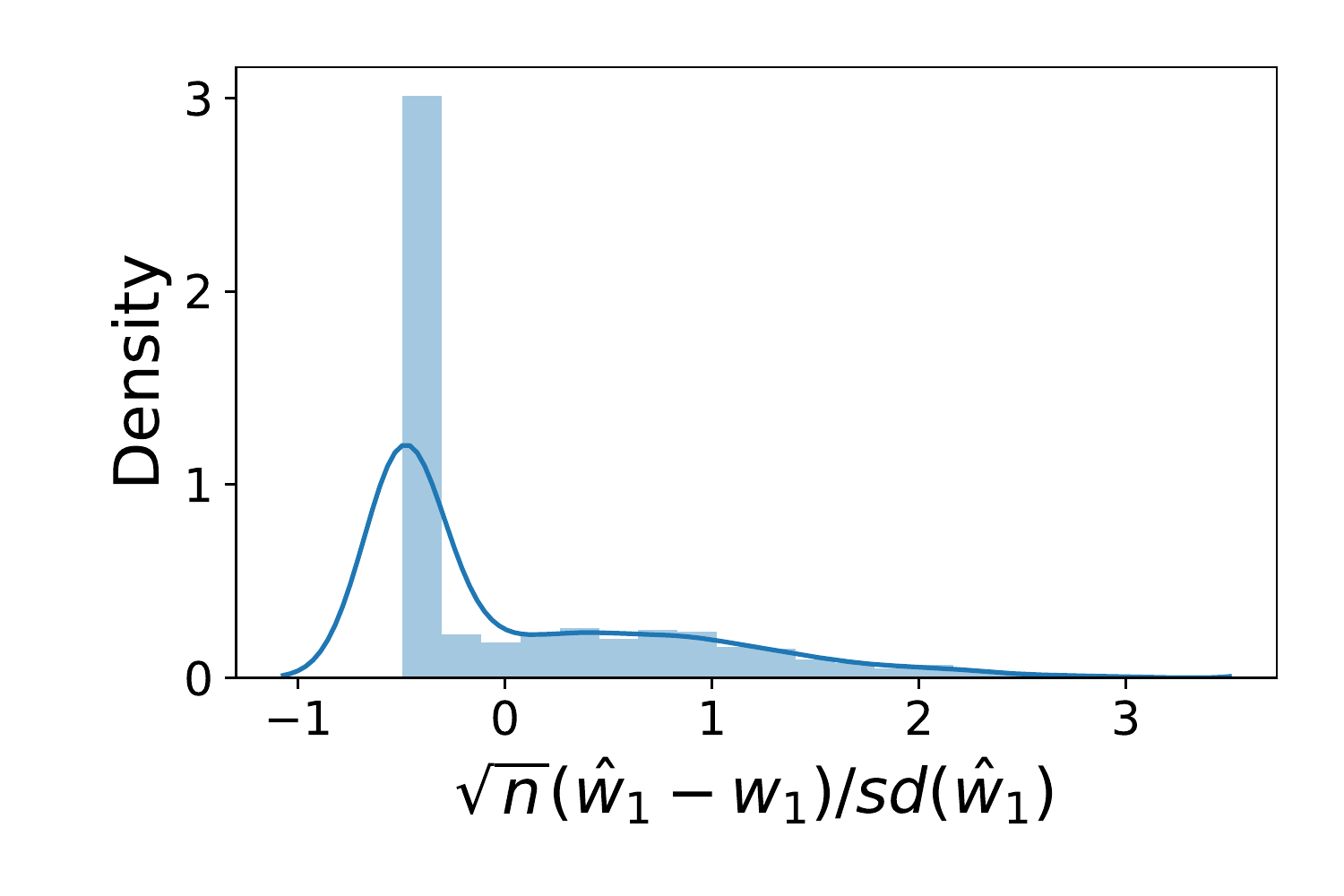}
                        \caption{Histogram of the large-sample distribution of the normalized $\hat{w}_1-w_1$ from EM estimator, which is not Gaussian.}
                        \label{fig:high_2}
                \end{minipage}
            \end{figure}

            As we showed in \Autoref{subsec:high-dim-mom}, the MoM with uniform design can be consistent in the high dimension case. Here we take $n=(p \ln p)^2$ (rounded) for $p \in \{4,6,8,10,12\}$ to mimic the case that $p$ grows with $n$ (\Autoref{thm_mom_dim}). 
            The reason that we did not choose a $p$ greater than $n$ is that a small $n$ tends to violate \Autoref{unique}. 
            As before, we generate a random population and an independence design for each replication. 
            We use $L_1$ loss $\norm{\widehat{\w}-\w}_1$ to assess the consistency. The mean losses from $k=1000$ replications are reported in \Autoref{fig:high_1}.
            All methods are consistent in the sense of $L_1$ loss. However, the asymptotic normality no longer holds for the MLE (as shown in  \Autoref{fig:high_2}).
            
        \subsubsection{Time cost}
\begin{figure}[tb]
                \begin{minipage}{0.45\textwidth}
                    \centering
                        \includegraphics[width=8cm]{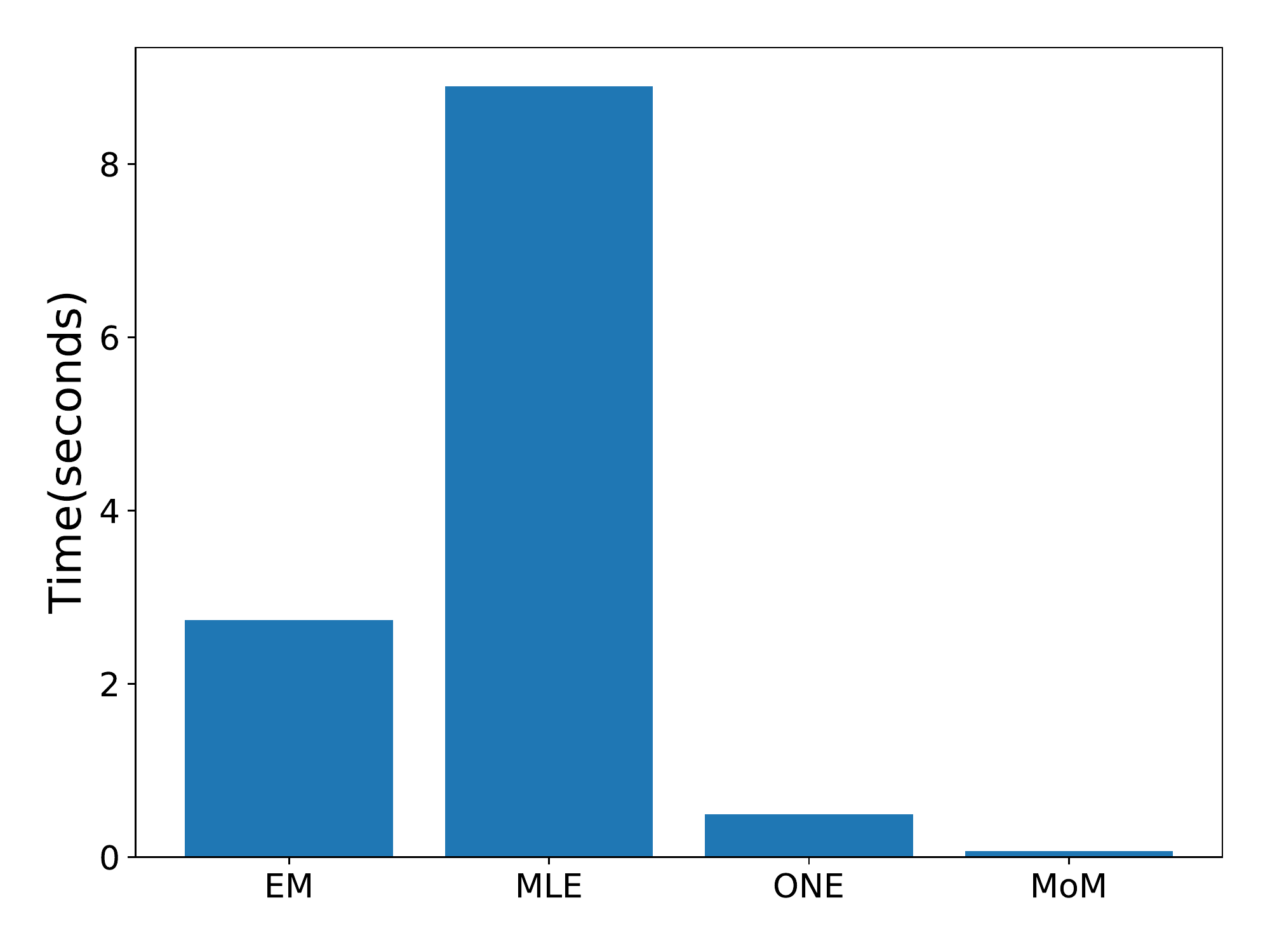}
                    \caption{Total time cost when $p=4, n=2000, k=1000$.}
                    \label{fig:time1}
                \end{minipage}
                \hfill
                \begin{minipage}{0.45\textwidth}
                    \centering
                        \includegraphics[width=8cm]{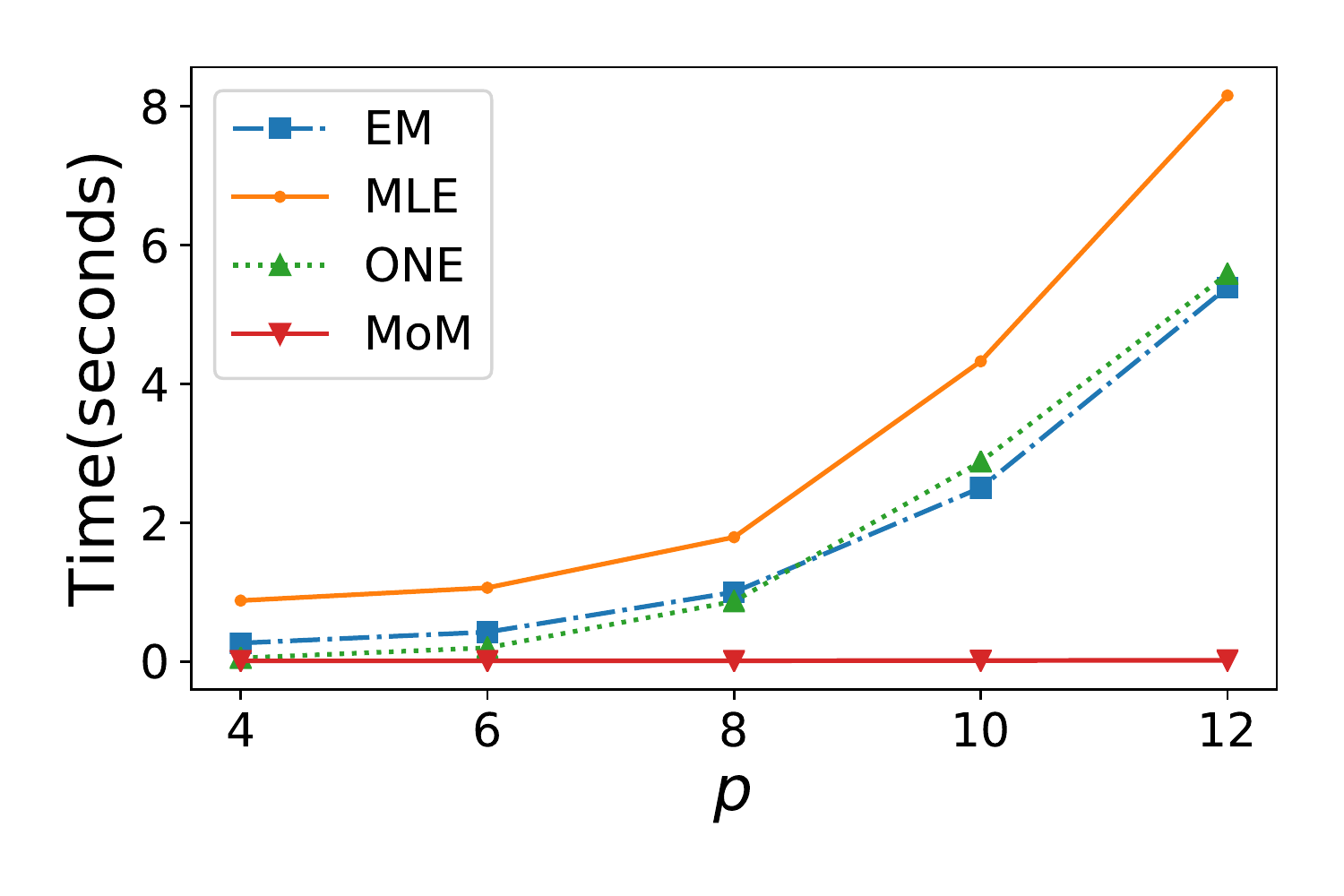}
                    \caption{Total time cost when $k=100, n=2000$, with $p$ varies.}
                    \label{fig:time2}
                \end{minipage}
            \end{figure}

            Using the same data-generating process, we report the total time cost of $k=100$ runs when $p=4,6,8,10,12$ for each method, with sample size $n=2000$. Additional comparison when $p=4, k=1000$ is given to highlight that MoM is significantly faster than other methods. The results are summarized in \Autoref{fig:time1, fig:time2}.
            
            We can see the MoM is the fastest as it only needs to solve a $p$-equation linear system. It's a well-understood optimization problem with time complexity no more than $O(p^3)$. Each iteration in EM is $O(np)$, and the number of iterations to converge will increase with $p$. The one-step estimator behaves similarly with EM when $p\geq8$. The time cost of one-step mainly comes from the calculation of the Hessian matrix, which is $O(np^2)$. The MLE by \textit{CVXPY} is the slowest. When $p$ is small, say $4$, MoM is already about $100$ times faster than EM, and $1000$ times faster than MLE by \textit{CVXPY}. 
            Overall, we can see that the asymptotic variance of MLE is a little bit smaller than MoM. But MoM is much faster than MLE. In practice, we recommend the use of MoM if we know the design, or the one-step estimator for a moderately large $p$.

    \subsection{Performance of independence tests} \label{subsec_indtest}
        As an important application of subset privacy, we perform independence tests described in \Autoref{sec_form} for contingency table from subset privacy and compare their performance. 
        Specifically, we consider the Pearson's Chi-square test (`Pearson'), likelihood ratio test designed for subset privacy (`LRT:MLE'), likelihood ratio test with the MLE approximated by MoM (``LRT:MoM''), Bonferroni correction-based test (`Bon').
        
\begin{figure}[tb]
                \begin{minipage}{0.45\textwidth}
                    \centering
                        \includegraphics[width=\textwidth]{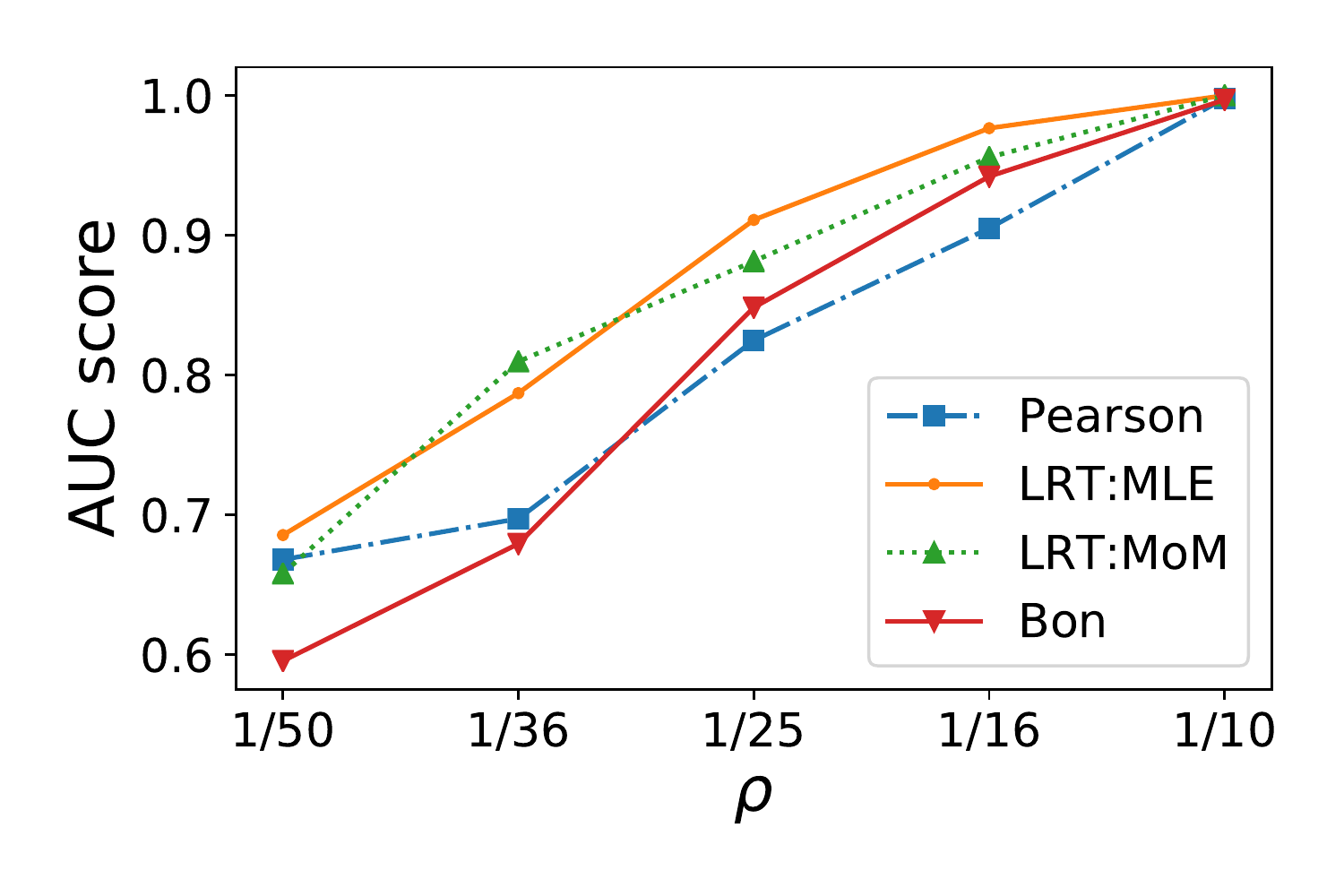}
                    \caption{AUC score of four tests when $\rho$ varies given $p=q=4, n=1000$}
                    \label{fig:auc-rho}
                \end{minipage}
                \hfill
                 \begin{minipage}{0.45\textwidth}
                    \centering
                        \includegraphics[width=\textwidth]{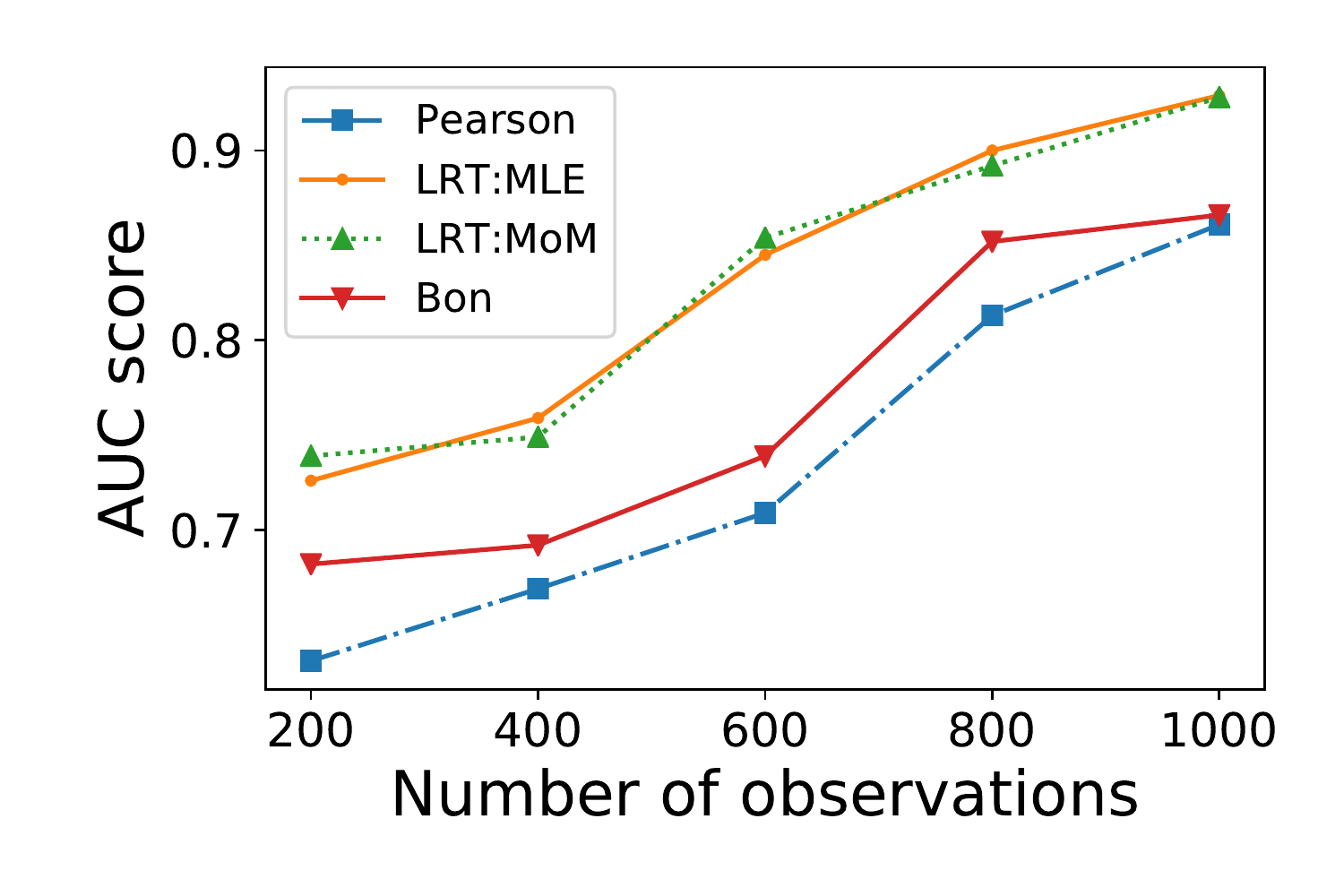}
                    \caption{AUC score of four tests when $n$ varies given $p=q=4, \rho=0.05$}
                    \label{fig:auc-n}
                \end{minipage}
            
            \end{figure}

\begin{figure}[tb]
                \begin{minipage}{0.45\textwidth}
                    \centering
                        \includegraphics[width=\textwidth]{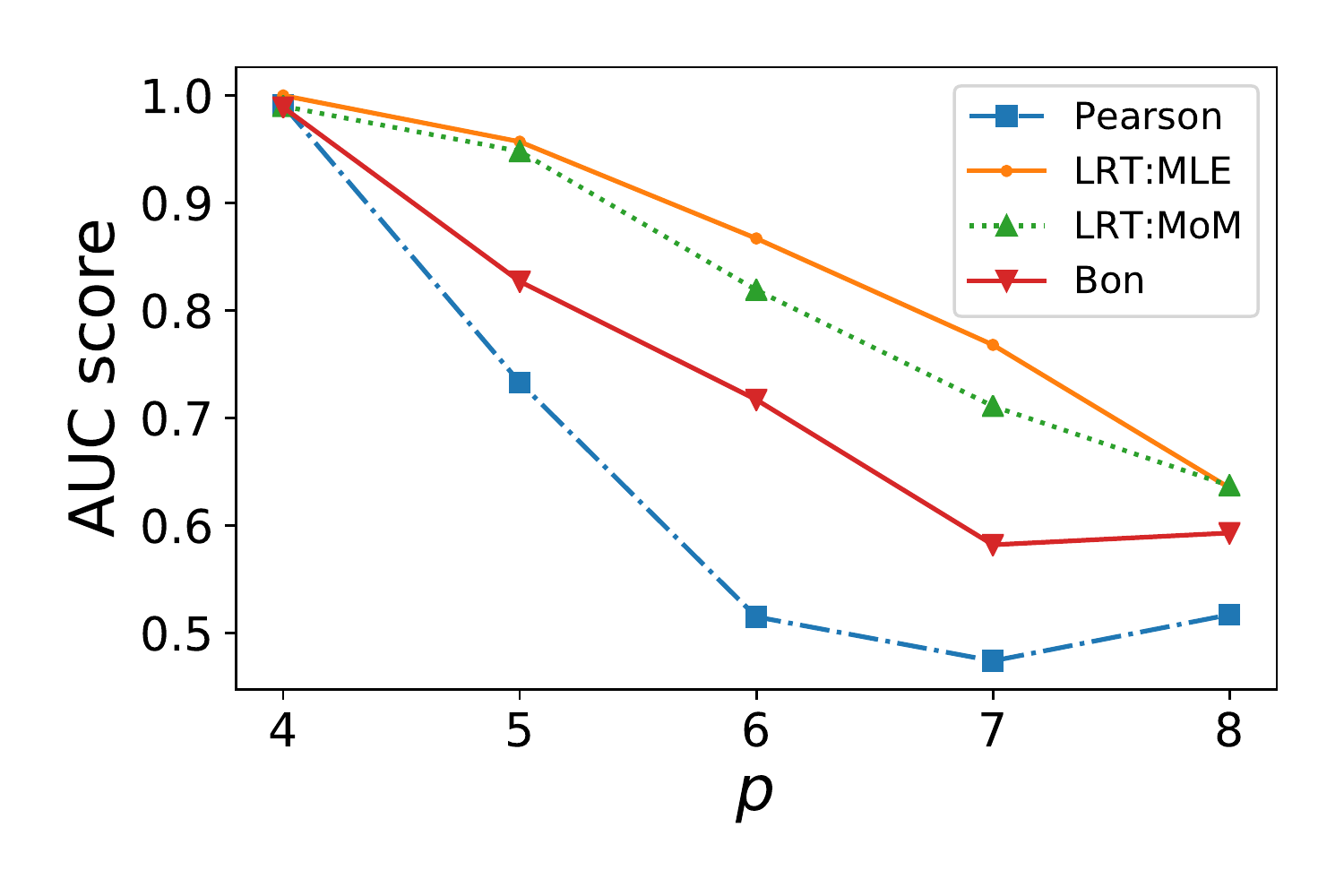}
                    \caption{AUC score of four tests when $p$ varies given $n=2000, \rho=0.05$}
                    \label{fig:auc-p}
                \end{minipage}
                \hfill
               \begin{minipage}{0.45\textwidth} \includegraphics[width=\textwidth]{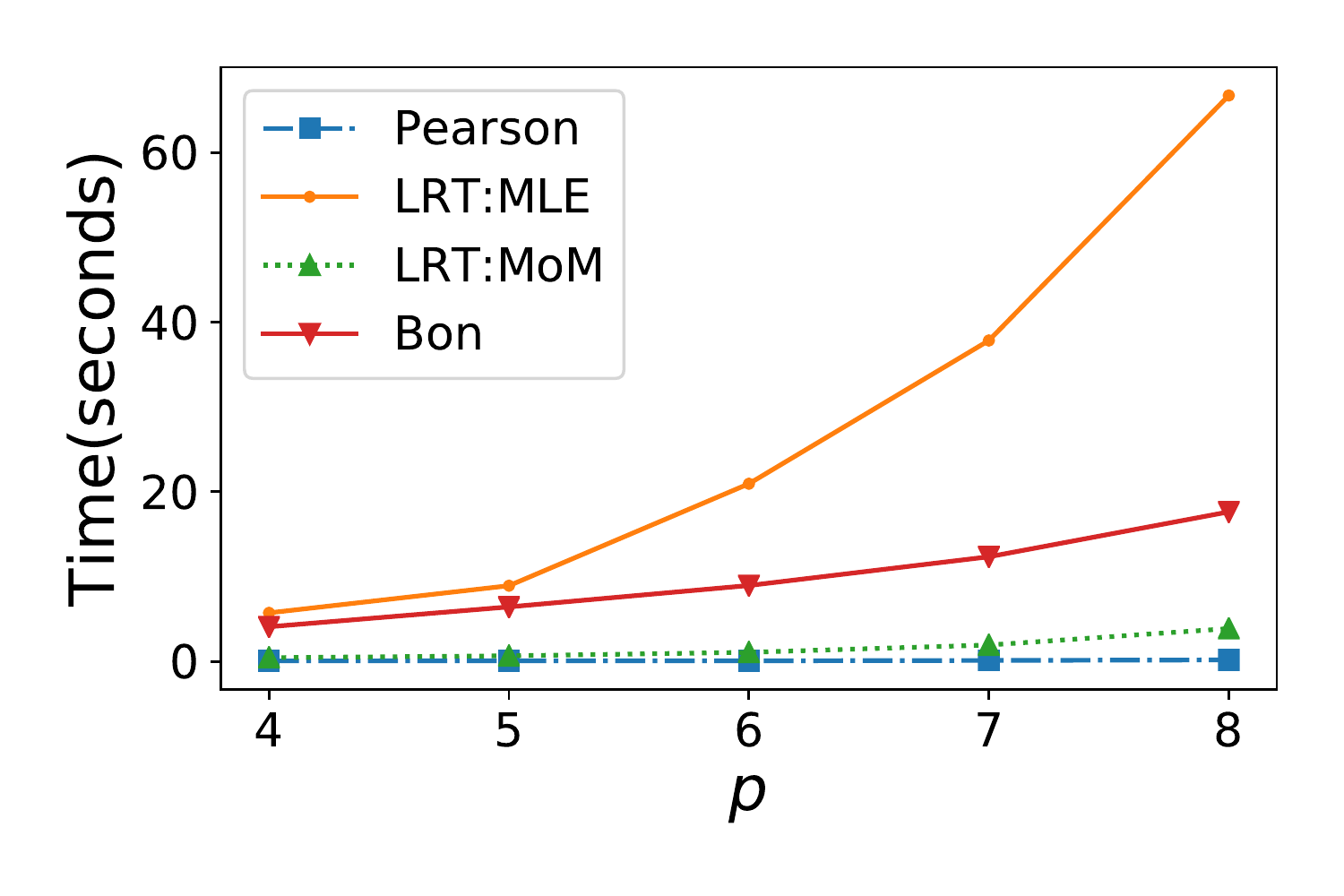}
                        \caption{Time cost for four tests, $n=2000, k=100$}
                        \label{fig:ct-time}
                \end{minipage}
               
            \end{figure}

        In the experiments, we choose $k=200$ replications and $p=q$ for simplicity. The data are generated as follows. The first $100$ replications are in an independent setting, and the remaining are in a dependent setting. In the independent setting, we randomly generate two independent random variables $X\sim \w_{X},Y\sim \w_{Y}$ and a product design. In the dependent setting, we generate true labels from a joint distribution $(X, Y) \sim W$, where $W \propto (\w_{X} \w_{Y}^\T + \rho I_p) $, and $\rho$ is a hyper-parameter that controls the level of dependence. A small value of $\rho$ means $X$ and $Y$ are almost independent. Then we sample subset observations according to the product design and apply the four tests to calculate \textit{p}-values. 
        To evaluate the performance, we use the area under the receiver operating characteristic curve (ROC-AUC).
        We inspect the change of ROC-AUC score against three factors.
        \begin{enumerate}
            \item $\rho$ varies given $p=q=4, n=1000$ (\Autoref{fig:auc-rho}).
            \item  $n$ varies given $p=q=4, \rho=0.05$ (\Autoref{fig:auc-n}).
            \item $p$ varies given $n=1000, \rho=0.05$ (\Autoref{fig:auc-p}).
        \end{enumerate}
        The above $\rho$ is chosen so that the signals are neither too weak nor strong in the test.
        The time cost is shown in \Autoref{fig:ct-time}. 

        For all the tests, the AUC score tends to decrease if $p$ increases, $n$ decreases, or $\rho$ (dependence) decreases, with other factors fixed. Among them, the LRT:MLE and LRT:MoM outperform the Pearson and Bon tests. Also, LRT:MoM has similar AUC scores compared with LRT:MLE, while the former's computation time is much shorter. From the experimental study, we recommend the use of LRT:MoM in practice. 
        
        \textit{Controlling the Type-I error in practice}.
            For real-world data, the distribution of the \textit{p}-value under the null hypothesis may not be close to the asymptotic result (if it exists). We suggest a practical way to control the Type-I error of the tests. Let $\alpha$ be our significance level. We calculate the \textit{p}-value under the null hypothesis and reject if the value is in the lower $\alpha$-quantile of the empirical distribution of the \textit{p}-value under the null.  
            Such an empirical distribution is obtained in the following way. 
            For subset samples $(A_i, B_i), i=1,\dots,n$, we randomly shuffle the order of $A_i$, so that $A$ and $B$ become approximately independent. 
            We repeat this procedure many times and calculate the \textit{p}-value for each replication, then use the results as the empirical distribution. 
            
\balance
\bibliography{privacy}
\bibliographystyle{IEEEtran}



\end{document}